\documentclass[12pt]{article}
\usepackage[utf8]{inputenc}
\usepackage[T1]{fontenc}

\usepackage[nocompress]{cite}
\usepackage{IEEEtrantools,stackengine}
\stackMath

\usepackage{authblk}
\usepackage{fullpage}
\usepackage{amssymb,amsmath}
\usepackage{float}
\usepackage{bm}
\usepackage{makecell}
\usepackage{siunitx}
\usepackage{csquotes}

\usepackage{xspace}
\newcommand\ie{i.e.\xspace}
\newcommand\eg{e.g.\xspace}

\newcommand\US{U.\,S.\xspace}

\newcommand{\var}[1]{\mathit{#1}}

\def\sym#1{\ifmmode^{#1}\else\(^{#1}\)\fi}
\usepackage{booktabs}
\usepackage{longtable}
\usepackage{multirow}
\usepackage[export]{adjustbox}

\usepackage[dvipsnames]{xcolor}

\usepackage[left]{lineno}

\nolinenumbers

\usepackage[
	colorinlistoftodos,
	textsize=footnotesize,
]{todonotes}

\definecolor{darkgreen}{rgb}{0.0, 0.5, 0.0}

\usepackage{setspace}
\usepackage[unicode=true]{hyperref}
\hypersetup{breaklinks=true,
	bookmarks=true,
	colorlinks=true,
	pdfborder={0 0 0},
	allcolors=blue}
\expandafter\def\expandafter\UrlBreaks\expandafter{\UrlBreaks
	\do\-}

\usepackage[%
	sort&compress
]{cleveref}
\Crefname{appendix}{Supplement}{Supplements}

\usepackage{times}
\usepackage{enumerate}
\usepackage{float}

\usepackage{subcaption}
\usepackage{graphicx}

\graphicspath{{./figures/}}

\usepackage[figuresright]{rotating}
\usepackage{tabularx}
\usepackage{dcolumn}
\usepackage{pdflscape}

\usepackage{array}

\newcolumntype{L}[1]{>{\raggedright\let\newline\\\arraybackslash\hspace{0pt}}p{#1}}
\newcolumntype{C}[1]{>{\centering\let\newline\\\arraybackslash\hspace{0pt}}p{#1}}
\newcolumntype{R}[1]{>{\raggedleft\let\newline\\\arraybackslash\hspace{0pt}}p{#1}}

\usepackage{sectsty}
\subsubsectionfont{\normalfont\itshape}

\makeatletter
\renewcommand{\fps@figure}{H}
\renewcommand{\fps@table}{H}
\makeatother

\allowdisplaybreaks

\usepackage{eurosym}

\usepackage{comment}

\usepackage{ntheorem}
\theoremseparator{:}

\usepackage[shortcuts]{extdash}
\usepackage{paralist}
\usepackage{colortbl} 
\usepackage{tabu}
\usepackage{bbm}
\usepackage{soul}

\DeclareSIUnit\week{week}
\DeclareSIUnit\weeks{weeks}
\DeclareSIUnit\billion{billion}

\begin{document}


\title{\centering\LARGE\singlespacing Community-based fact-checking reduces the spread of misleading posts on social media}

\renewcommand\Affilfont{\fontsize{9}{10.8}\selectfont}

\author[1]{Yuwei Chuai$^\dagger$}
\author[2]{Moritz Pilarski$^\dagger$}
\author[3]{Thomas Renault$^\dagger$}
\author[4]{David Restrepo-Amariles}
\author[4]{Aurore Troussel-Clément}
\author[1]{Gabriele Lenzini}
\author[2]{Nicolas Pröllochs}

\affil[1]{University of Luxembourg, Luxembourg}
\affil[2]{JLU Giessen, Germany}
\affil[3]{Université Paris 1 Panthéon-Sorbonne, France}
\affil[4]{HEC Paris, France}

\date{}

\maketitle

$^\dagger$These authors contributed equally to this work.

\clearpage
\begin{abstract}
\normalfont
\noindent
Community-based fact-checking is a promising approach to verify social media content and correct misleading posts at scale. Yet, causal evidence regarding its effectiveness in reducing the spread of misinformation on social media is missing. Here, we performed a large-scale empirical study to analyze whether community notes reduce the spread of misleading posts on X. Using a Difference-in-Differences design and repost time series data for $N=$~\num{237677} (community fact-checked) cascades that had been reposted more than \num{431} million times, we found that exposing users to community notes reduced the spread of misleading posts by, on average, 62.0\%. Furthermore, community notes increased the odds that users delete their misleading posts by \num{103.4}\%. However, our findings also suggest that community notes might be too slow to intervene in the early (and most viral) stage of the diffusion. Our work offers important implications to enhance the effectiveness of community-based fact-checking approaches on social media. 
\end{abstract}

\flushbottom
\maketitle
\thispagestyle{empty}


\sloppy
\raggedbottom


\clearpage
\section*{Introduction}
\label{sec:introduction}


The spread of misinformation on social media platforms has become a concerning issue of the digital age and raised alarm bells across various domains. Negative repercussions of misinformation have been repeatedly observed, with tangible consequences in critical areas such as elections \cite{Allcott.2017,Aral.2019,Bakshy.2015,Grinberg.2019,Guess.2020,Moore.2023}, public health \cite{Broniatowski.2018,Rocha.2021,Gallotti.2020,Roozenbeek.2020}, and public safety \cite{Starbird.2017,Bar.2023,Oh.2013}. Researchers, governments, and regulation authorities thus urge social media providers (\eg, X/Twitter, Facebook) to develop effective countermeasures to counteract the spread of misinformation on their platforms \cite{Lazer.2018,Calo.2021,Donovan.2020,Kozyreva.2022}.


To this end, a widely implemented approach is the use of professional fact-checkers to identify and label misleading posts \cite{Facebook.2016,Instagram.2019}. The rationale is that if users are warned that a message is false, they should be less likely to believe it. While this approach has been shown to be effective in numerous experimental studies {\cite{Altay.2023,Kim.2019,Ng.2021,Pennycook.2020b,Porter.2021,Moravec.2020,Pennycook.2020,Clayton.2020,Mena.2020,Yaqub.2020,Martel.2023b} (for a review, see \cite{Martel.2023b}), it faces several key challenges, particularly with regard to volume, visibility, and trust. First, in\-/depth investigation of claims can be time\-/consuming, often taking many hours or even days \cite{Guo.2022}. Given the limited number of available fact\-/checkers, not every claim can be thoroughly examined. Consequently, fact\-/checkers are compelled to prioritize content that is overtly false or intentionally misleading over material that is more intricate or nuanced \cite{Pennycook.2019}. Second, professional fact\-/checks frequently have a restricted audience. Except for some collaborations with social media platforms on particular topics, fact\-/checking organizations primarily disseminate their findings through their own websites. A study conducted in 2017 revealed that over half of all \US adults had never visited any fact\-/checking website \cite{Robertson.2020}, demonstrating the limited reach of professional fact\-/checks. Third, a major drawback is that professional fact\-/checkers face a lack of trust from large portions of society \cite{Straub.2022,Poynter.2019}. For example, according to surveys, the majority of Republican partisans (\SI{70}{\percent}) and half of all \US adults believe that fact\-/checkers exhibit bias \cite{Poynter.2019}.


As a remedy, research has proposed to employ non-expert fact-checkers in the crowd to fact-check social media content \cite{Micallef.2020,Bhuiyan.2020,Pennycook.2019,Epstein.2020,Allen.2020,Allen.2021,Godel.2021,Martel.2023}. This approach is based on the idea that the biases and errors of a single user can be mitigated by tapping into the collective intelligence of diverse individuals, also known as the ``wisdom of the crowds'' \cite{Frey.2021,Woolley.2010,Martel.2023}. Experimental studies found that the assessments of even relatively small crowds are comparable to those of experts \cite{Bhuiyan.2020,Epstein.2020,Pennycook.2019,Martel.2023,Godel.2021,Allen.2021,Resnick.2021}. Furthermore, crowd-based fact-checking enables the examination of significantly larger volumes and varieties of posts as compared to expert-based assessments \cite{Prollochs.2022a,Drolsbach.2023b,Pilarski.2024,Zhao.2020}. Also, community notes mitigates trust issues with simple misinformation flags \cite{Drolsbach.2024}. 
Altogether, community-based fact-checking may have the potential to address many of the drawbacks of expert-based fact-checking on social media.


Building upon these encouraging findings, the social media platform X (formerly Twitter) has recently introduced ``Community Notes'' (formerly ``Birdwatch''), a community-based fact-checking system that allows users to assess the accuracy of posts \cite{Prollochs.2022a, Twitter.2021}. ``Community Notes'' is the first large-scale attempt to implement community-based fact-checking on a major social media platform. It enables enrolled users to flag posts that they believe are misleading and add concise fact-checking explanations, which are then rated by other users for their helpfulness. Notes rated by the community to be helpful then appear directly on the fact-checked misleading post to inform other users. However, empirical evidence on whether displaying community notes reduces the spread of misleading posts on social media is missing so far.


In prior research, the efficacy of fact-checking has mainly been studied in one of two forms: (i) lab/survey experiments evaluating whether flagging misinformation influences misinformation discernment and sharing intentions \cite{Kim.2019,Ng.2021,Pennycook.2020b, Moravec.2020,Pennycook.2020,Clayton.2020,Mena.2020}, and (ii) field studies analyzing the spread of misleading vs. non-misleading posts \cite{DelVicario.2016,Vosoughi.2018,Chuai.2024,Friggeri.2014, Drolsbach.2023b}. The former group of studies, \ie, lab/survey experiments, have shown that misinformation flags are effective in increasing the participants' ability to discern between true and false content \cite{Pennycook.2020b, Moravec.2020, Clayton.2020} as well as decreasing their sharing intentions for misinformation \cite{Kim.2019, Mena.2020, Ng.2021}. The latter of studies, \ie, field studies, were largely limited to correlational works analyzing the virality of misleading vs. non-misleading posts on social media \cite{Vosoughi.2018,Drolsbach.2023b,Chuai.2024,Friggeri.2014,DelVicario.2016}. As an example, Vosoughi et al. (2018) \cite{Vosoughi.2018} found that expert fact-checked misinformation is linked to more viral resharing cascades. Still, these works have not quantified the real-world intervention effect of fact-checking, mainly due to data constraints such as missing time series data on reposts and interventions. Here, we add by analyzing the efficacy of community notes in reducing the spread of misleading posts on the social media platform X. 


In this work, we performed a large-scale quasi-experimental study to analyze whether community fact-checks reduce the spread of misleading posts on the social media platform X. For this, we examined $N=\,$\num{237677} (community fact-checked) X cascades that were created within a period of over 20 months from the roll-out of ``Community Notes'' on October 6, 2022 to June 11, 2024, and had been reposted more than \num{431} million times. To evaluate the efficacy of community notes in reducing the spread of misleading posts on X, we collected \emph{time series data} of repost counts for the community fact-checked posts over 36 hours since their creation. We then employed a Difference-in-Differences (DiD) design and negative binomial regression models to estimate the reduction in reposts for misleading posts after the display of community notes. Overall, this setting offers a unique opportunity to study the real-world efficacy of community-based fact-checking on a major social media platform and enables us to quantify the intervention effect of community notes.


As a secondary analysis, we examined the extent to which the ``Community Notes'' feature in its current implementation reduced the spread of misleading posts on X in terms of cumulative repost count. The underlying rationale is that the display timing of the notes might be critical. Even if users are less inclined to share tweets once they are equipped with a community note, the fact-check might arrive too late to intervene in the early and most viral stage of the diffusion, thereby potentially diminishing its effect on overall engagement with misinformation on social media. 

\clearpage
\section*{Results}
\label{sec:results}

\subsection*{Data Overview}


The ``Community Notes'' program on X enables users to flag posts that they believe are misleading and contribute textual notes that provide context to the source post (see example in \Cref{fig:data_overview}a). After a community note is submitted, the fact-check becomes available for other enrolled contributors to rate while remaining hidden from the public users. To surface helpful community notes that appeal broadly across heterogeneous user groups, the program features a bridging-based rating system \cite{Wojcik.2022}. This system calculates the helpfulness score for each community note based on the ratings made by the contributors. Only notes that are rated as helpful by multiple contributors with heterogeneous rating histories are displayed to public users on X \cite{Twitter.2024b}.


To evaluate the efficacy of community notes in reducing the spread of misleading posts on X, we gathered time series data on repost counts and fact-checking histories for \num{237677} posts that have been fact-checked on X's ``Community Notes'' platform (see \nameref{sec:methods}). Our dataset covers all community fact-checked posts written in English between the roll-out of ``Community Notes'' to the general public on October 6, 2022, and June 11, 2024, \ie, for an observation period of over 20 months. For each post, we used the X API v2 to retrieve the time series of repost histories at minute level over 36 hours after its creation, during which the vast majority of all reposts occured \cite{Chuai.2022,Pfeffer.2023}. Throughout our observation period, the posts in our dataset had been reposted more than \num{431} million times. Subsequently, we combined the time series data on repost counts with fact-checking histories indicating at which time point the community notes were rated helpful and displayed to the public users. The data collection process resulted in a longitudinal dataset comprising information about the repost counts at minute level and the current note status for each fact-checked post (see \Cref{supp:descriptives} for summary statistics). This allows us to scrutinize the efficacy of community notes in reducing the spread of misleading posts on X.


\Cref{fig:data_overview}b shows that the ``Community Notes'' feature gained significant traction after its roll-out to the public on October 6, 2022. Both the number of community notes and fact-checked source posts increased significantly. On average, community note contributors fact-checked \num{386} posts in total per day. Here, multiple users can write notes for the same post. Therefore, the data sometimes includes multiple fact-checks for the same post. The average number of notes per fact-checked post was \num{1.869}. Only a relatively small fraction of 10.1\% of all notes was actually displayed to users, \ie, were surfaced as helpful by the rating system of ``Community Notes.'' However, once community notes received the helpful status, they tended to maintain displayed stably on the fact-checked posts (see \Cref{supp:note_status}).


Compared to the rapid dissemination of posts on X, fact-checking via community notes was relatively slow. While 75.9\% of all notes deemed helpful were displayed to general users within 36 hours after their creation, the time lag between post creation and note display (\ie, the response time) averaged at \num{61.4} hours (median of \num{18.1} hours; see \Cref{fig:data_overview}c). In contrast, posts on X spread considerably faster: the average half-life of posts (\ie, the post age at which the cumulative ratio of reposts reaches to 50\% of 36-hour reposts) amounted to merely \num{6.25} hours for posts with displayed notes  (\Cref{fig:data_overview}d). Thus, community notes were unlikely to have a meaningful effect on the cumulative repost count. For posts without displayed notes, the average half-life of posts is \num{5.75} hours, which is slightly but significantly ($\var{t} = \num{13.036}, p < \num{0.001}$) smaller than for posts with displayed notes. \Cref{fig:data_overview}e further shows that source posts with displayed notes tended to receive a higher cumulative number of 36-hour reposts (median of \num{432}) on average, compared to those that did not have any displayed notes (median of \num{258}, $\var{KS} = \num{0.117}, p < \num{0.001}$). Overall, these observations support earlier findings \cite{Chuai.2024}, indicating that community notes might be too slow to curb overall engagement with misleading posts on X. Notwithstanding, even if community notes might be too slow to have a meaningful effect at the cumulative level, it is still entirely possible (and plausible) that they reduced reposts for posts with displayed notes relative to similar posts without displayed notes. Quantifying this intervention effect presents the main objective of our study. 

The data further suggests that posts with displayed notes were more likely to be deleted on X (see \Cref{fig:data_overview}e). On average, the deletion ratio was 17.3\% for posts with displayed notes and 9.8\% for posts without displayed notes.

\begin{figure}
	\captionsetup[subfloat]{font={bf, small}, skip=0pt, singlelinecheck=false, labelformat=simple, position=top}
	\centering
	\begin{minipage}{\textwidth}
		\centering
		\begin{minipage}{0.34\textwidth}
		\subfloat[]{\fbox{\includegraphics[width = 0.9\textwidth, keepaspectratio]{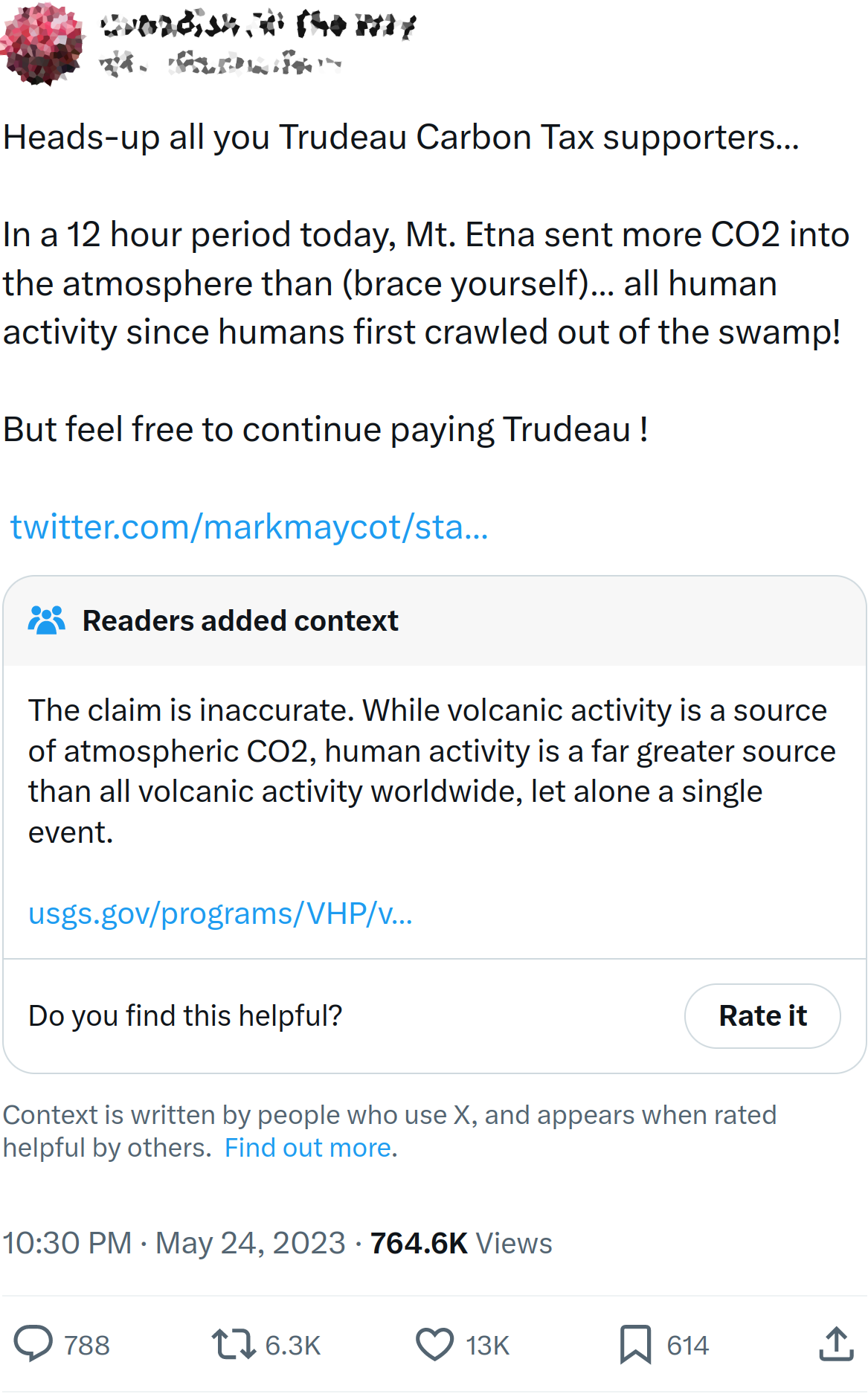}}}
		\end{minipage}
		\hfill
		\begin{minipage}{0.65\textwidth}
			\subfloat[]{\includegraphics[width = .49\textwidth]{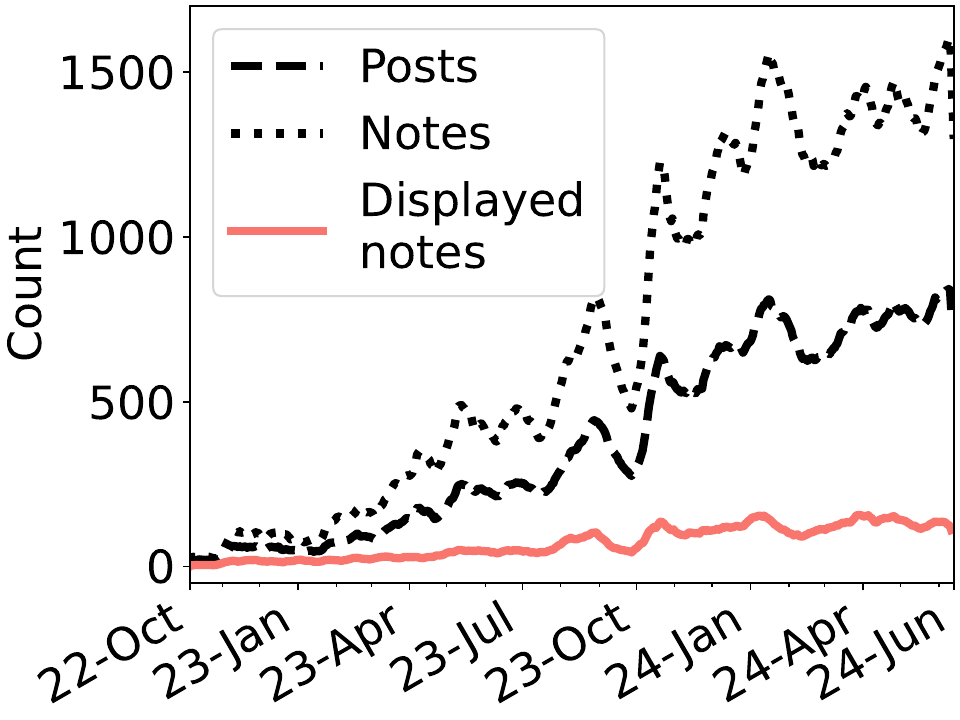}}
			\hfill
			\subfloat[]{\includegraphics[width = .49\textwidth]{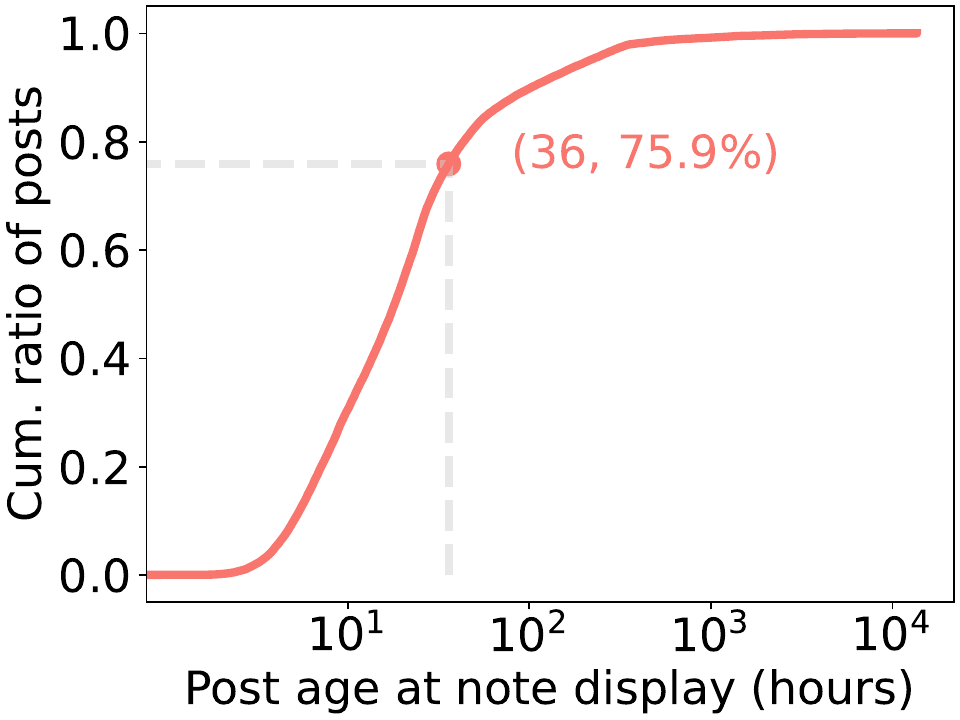}}
            \\
			\subfloat[]{\includegraphics[width = .49\textwidth]{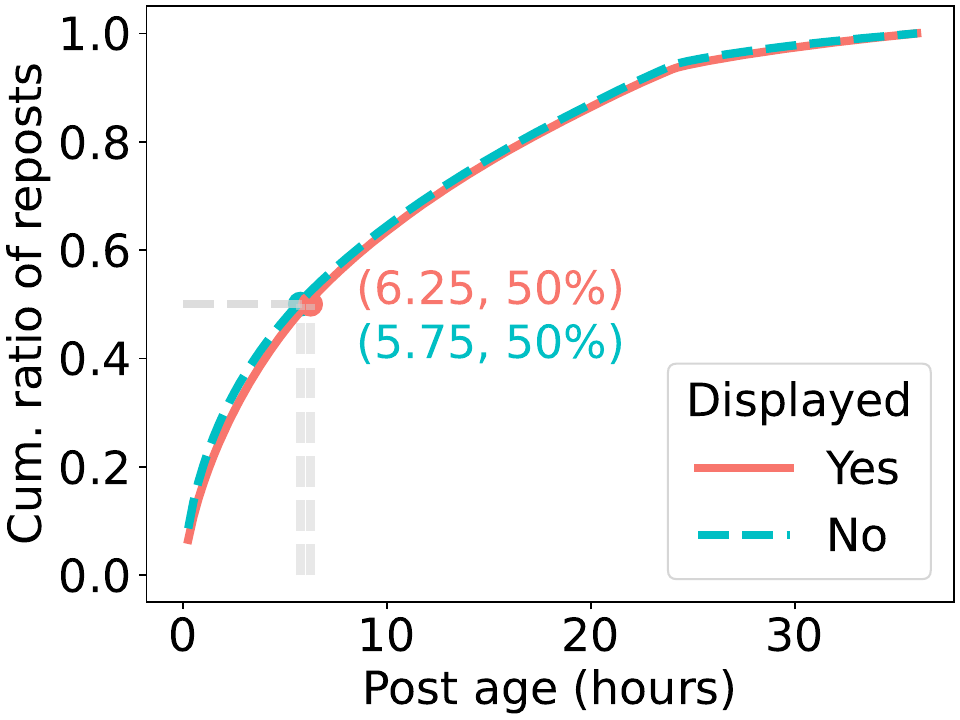}}
            \hfill
            \subfloat[]{\includegraphics[width = .49\textwidth]{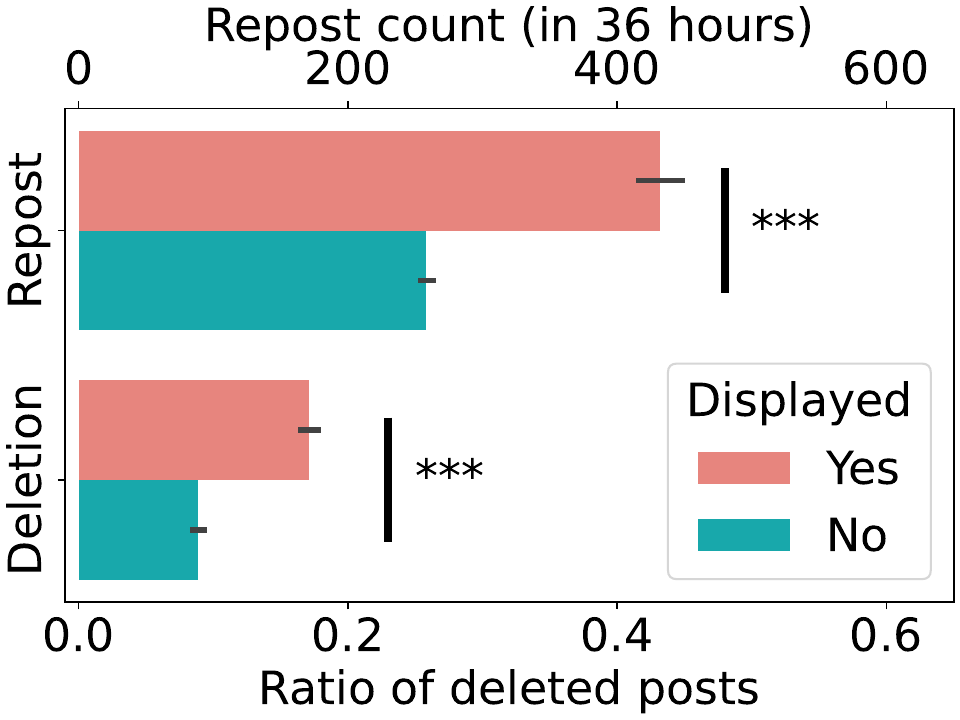}}
		\end{minipage}
	\end{minipage}
	\caption{\textbf{Data overview.} \textbf{(a)}~An example of a community note displayed on a misleading post on X. \textbf{(b)}~Two-week rolling averages of the daily counts of fact-checked source posts, community notes, and displayed notes in our observation period from October 6, 2022 to June 11, 2024. \textbf{(c)}~The cumulative distribution of the ratios of posts that received displayed notes at different post ages relative to all posts with displayed notes. \textbf{(d)}~The cumulative distribution of the ratios of reposts at different post ages relative to all reposts within 36 hours. \textbf{(e)}~The repost counts of retrieved posts (upper half plot, shown are median values) and the ratios of deleted posts (bottom half plot, shown are mean values) within groups of posts with displayed notes and posts without displayed notes. The error bars represent 99\% Confidence Intervals (CIs).}
	\label{fig:data_overview}
\end{figure}

\newpage
\subsection*{Difference-in-Differences Estimation}


We employed a Difference-in-Differences (DiD) design and negative binomial regression models to estimate the Average Treatment effect on the Treated (ATT), \ie, the extra repost reduction in the treatment group after the display of community notes compared to the baseline before the display and relative to the control group (see \nameref{sec:methods}). To mitigate potential confounding factors and ensure parallel trend, we constructed a control group from the source posts without displayed notes through one-to-one matching. The one-to-one matching was conducted based on the variables from user profiles (\eg, followers and followees) and post features (\eg, sentiments and topics). As a result, the source posts in the control group have no statistically significant difference with the source posts in the treatment group with respect to the user and post characteristics (see details in \Cref{supp:group_construction}). Subsequently, we assigned the virtual time of note display to the source posts in the control group according to the corresponding posts in the treatment and recenter the repost timelines of the source posts around the display of community notes in the treatment and control groups.


We started our analysis with a two-period DiD model. This allowed us to estimate difference in outcomes (\ie, repost counts) between the treatment and control group before and after the treatment (\ie, displaying community notes) was applied. To this end, we considered the period between 1 and 12 hours from the note display as the after-display period  and the period between 2 to 4 hours before the note display as the before display period (see \nameref{sec:methods}). The ATT for the two-period DiD model is visualized in purple color in \Cref{fig:regression_results} (see \Cref{supp:estimation_results} for full estimation results). The estimate for the ATT was \num{-0.620} (99\% CI: [\num{-0.625}, \num{-0.615}]; $p<\num{0.001}$), which indicates that displaying community notes reduced the subsequent number of reposts by, on average, 62.0\%. This suggests a pronounced treatment effects of community notes in reducing the spread of misleading posts on X.


Subsequently, we estimated a multi-period DiD to enable a temporal analysis of the treatment effect. Here, we again considered a 4-hour window before the note display as then before-display period and then examined the hourly multi-period ATTs from 1 to 12 hours after the note display (see \nameref{sec:methods}). \Cref{fig:regression_results} visualizes the hourly multi-period ATTs over time (see \Cref{supp:estimation_results} for full estimation results). We observed that during the hours before the note display, the ATT estimates were not statistically significantly different from zero (each $p > 0.001$), which supports the parallel assumption. However, we observed statistically significant ATTs after the display of community notes. During the first one hour after the note display, the ATT was estimated as \num{-0.386} (99\% CI: [\num{-0.402}, \num{-0.369}]; $p<\num{0.001}$). This implies that community notes reduced the number of reposts by 38.6\% within the first hour. Furthermore, the efficacy of community notes increased over time following the note display. The ATTs were \num{-0.548} (99\% CI: [\num{-0.560}, \num{-0.536}]; $p<\num{0.001}$) at the second hour, \num{-0.629} (99\% CI: [\num{-0.639}, \num{-0.619}]; $p<\num{0.001}$) at the fourth hour, and \num{-0.659} (99\% CI: [\num{-0.668}, \num{-0.649}]; $p<\num{0.001}$) at the eighth hour. At the twelfth hour after the note display, the hourly ATT was \num{-0.682} (99\% CI: [\num{-0.691}, \num{-0.673}]; $p<\num{0.001}$). In sum, our temporal analysis indicates that community notes reduced the number of reposts from 38.6\% to 68.2\% within the first 12 hours after they were displayed to users.


Additionally, to ensure that the estimated ATTs are robust and attributed to the display of community notes, we undertook a placebo test  (see \Cref{supp:placebo}). Specifically, we replaced the treatment group with a placebo group. The results reveal that there was no additional reduction in reposts within the placebo group, which supports that the estimated ATTs in the treatment group can be attributed to the display of community notes.

\begin{figure}
	\captionsetup[subfloat]{font={bf, small}, skip=0pt, singlelinecheck=false, labelformat=simple, position=top}
	\centering
    \subfloat[]{\includegraphics[width = \textwidth]{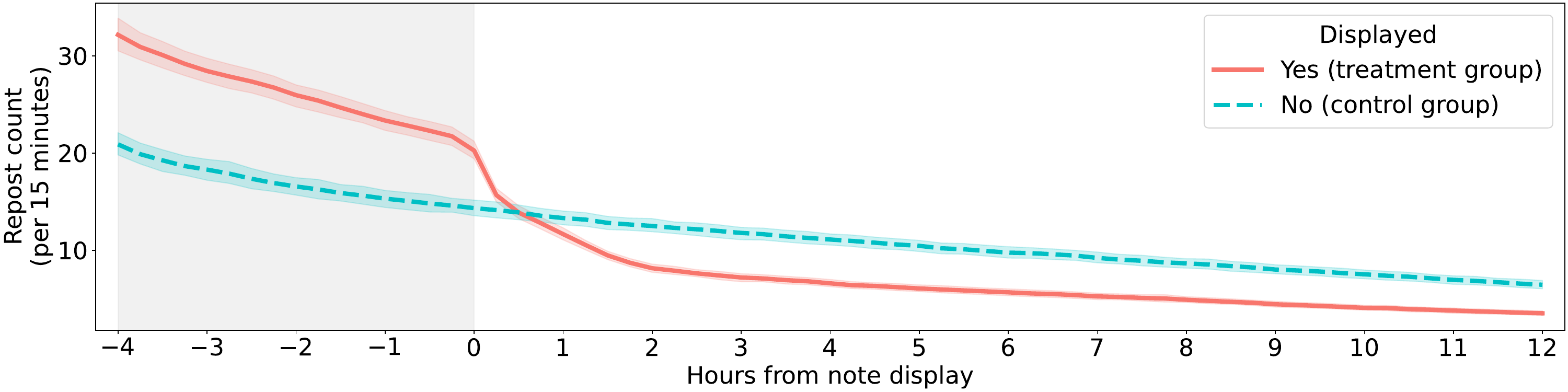}}
    \\
    \subfloat[]{\includegraphics[width = \textwidth]{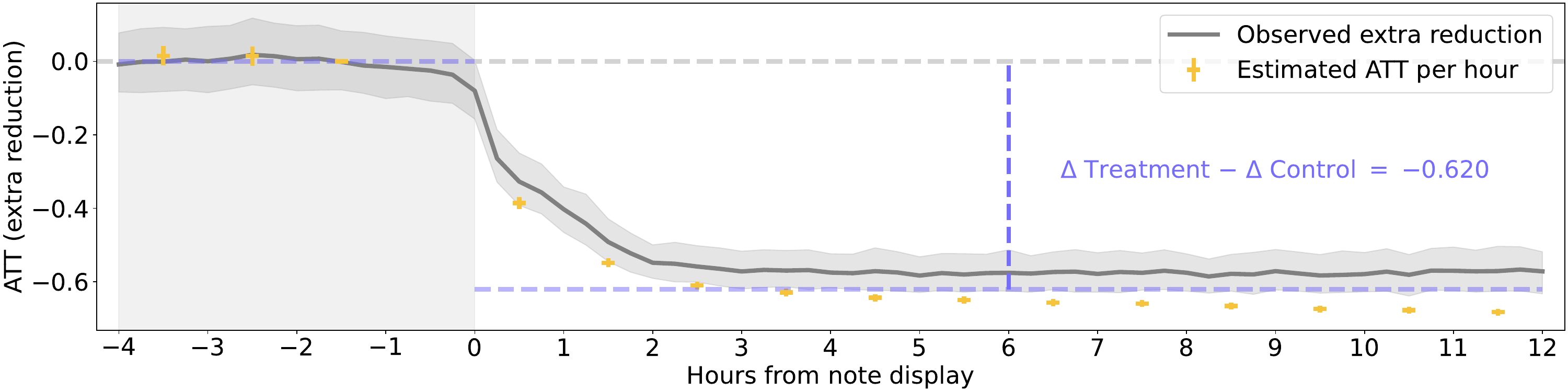}}
	\caption{\textbf{Community notes reduce the spread of misleading posts on X.} \textbf{(a)}~Time series of repost counts within 15 minutes intervals in the treatment group (red) and the control group (blue) from 4 hours before the display of community notes to 12 hours after the display of community notes. The error bands represent 99\% Confidence Intervals (CIs). \textbf{(b)}~Two-period (purple) and multi-period (yellow) ATTs estimated using a Difference-in-Differences (DiD) design and negative binomial regression models. For the two-period DiD model, the ATT is calculated as $\Delta$ Treatment $-$ $\Delta$ Control $=$ \num{-0.620}. For the multi-period DiD model, the yellow circles and error bars show the estimated hourly multi-period ATTs (with 99\% CIs). The grey band (with 99\% CIs) visualizes the observed extra reduction of the ratio of reposts in the treatment group relative to reposts in the control group and compared to the ratio of reposts before the display of community notes. The ATT estimations are based on \num{614520} repost time series observations for $N =$~\num{40968} posts. Post-level random effects are included. Full estimation results are in \Cref{tab:did_main_fixed,tab:did_main_multi}.}
	\label{fig:regression_results}
\end{figure}

\newpage
\subsection*{Sensitivity Analyses}


We conducted additional sensitivity analyses on how the efficacy of community notes varied (i) depending on the post age at note display, (ii) over time since the roll-out of the ``Community Notes'' program in October 2022, (iii) depending on the number of helpfulness ratings from other fact-checking contributors, and (iv) across user and post characteristics (see \Cref{supp:sensitivity}).


We started by analyzing the efficacy of community notes varies for early vs. late fact-checks, \ie, the sensitivity across different post ages at the time of note display. The estimated ATTs of community notes across subgroups separated by the post age at which the community notes were displayed are visualized in \Cref{fig:sensitivity}a. We found that the treatment effects were significantly more pronounced for early vs. late community notes. For instance, for community notes displayed within 4 to 8 hours since the creation of the misleading post, the ATT was \num{-0.680} (99\% CI: [\num{-0.688}, \num{-0.673}]; $p<\num{0.001}$). In contrast, the ATT decreased to \num{-0.461} (99\% CI: [\num{-0.479}, \num{-0.442}]; $p<\num{0.001}$) if the community notes was displayed within 20 to 24 hours after post creation. Overall, the efficacy of community notes in reducing the spread of misleading posts was higher if they were displayed earlier.


Next, we examined the changes in the efficacy of community notes since the roll-out of the feature in October 2022. The estimated ATTs of community notes across subgroups separated by the months in which the posts were created since the roll-out are visualized in \Cref{fig:sensitivity}b. We found that the ATTs in the first month and second month of the community notes program were only \num{-0.245} (99\% CI: [\num{-0.354}, \num{-0.116}]; $p<\num{0.001}$) and \num{-0.248} (99\% CI: [\num{-0.333}, \num{-0.152}]; $p<\num{0.001}$), respectively. However, after 21 months, the ATT increased to \num{-0.705} (99\% CI: [\num{-0.727}, \num{-0.681}]; $p<\num{0.001}$). Overall, this suggests an trend of increasing efficacy of community notes in the months following its launch. 


We also analyzed how the efficacy of community notes varies depending on the number of ratings, \ie, how it was perceived by other fact-checking contributors. To this end, we only considered source posts with community notes that indicate potentially misleading information and had never been rated as not helpful. Thus, a higher number of ratings points towards potential high-quality notes that attract a lot of attention from other fact-checking contributors (\eg, because the fact-checked post is particularly viral). \Cref{fig:sensitivity}c shows the ATT estimates across the rating thresholds from 10 to 80 (mean of ratings per note) with a step size of 10. We found that the ATT estimates across the rating thresholds had no significant differences with each other. This suggests that once the number of helpfulness ratings exceeded the display threshold, the efficacy of community notes in reducing the spread of misleading posts was relatively stable. 


Additionally, we studied whether the treatment effect of community notes was moderated by author characteristics (\eg, the number of followers, verified status) and post characteristics (\eg, sentiment, topics). \Cref{fig:sensitivity}c shows that the ATT estimates were significantly negative across all author characteristics (full results are in \Cref{tab:did_sensitivity_user_charcs}~--~\Cref{tab:did_sensitivity_post_charcs2}, \Cref{supp:sensitivity}). However, the efficacy of community notes was higher if the author of the fact-checked post was not verified (ATT: \num{-0.738}; 99\% CI: [\num{-0.747}, \num{-0.728}]; $p<\num{0.001}$), for younger accounts (ATT: \num{-0.654}; 99\% CI: [\num{-0.660}, \num{-0.647}]; $p<\num{0.001}$), and for accounts with a lower number of followers (ATT: \num{-0.692}; 99\% CI: [\num{-0.699}, \num{-0.686}]; $p<\num{0.001}$), as compared to fact-checked post from verified accounts (ATT: \num{-0.589}; 99\% CI: [\num{-0.594}, \num{-0.583}]; $p<\num{0.001}$), older accounts (ATT: \num{-0.584}; 99\% CI: [\num{-0.591}, \num{-0.577}]; $p<\num{0.001}$), and accounts with higher number of followers (ATT: \num{-0.558}; 99\% CI: [\num{-0.565}, \num{-0.551}]; $p<\num{0.001}$). This suggests that the effectiveness of community notes might be slightly discounted for source posts from accounts with high social influence.


In terms of post characteristics, the ATT estimates were higher for posts that were shorter (ATT: \num{-0.649}; 99\% CI: [\num{-0.655}, \num{-0.642}]; $p<\num{0.001}$) and posts that had media elements (ATT: \num{-0.644}; 99\% CI: [\num{-0.649}, \num{-0.638}]; $p<\num{0.001}$), as compared to posts that were longer (ATT: \num{-0.591}; 99\% CI: [\num{-0.598}, \num{-0.583}]; $p<\num{0.001}$) and posts without media elements (ATT: \num{-0.539}; 99\% CI: [\num{-0.551}, \num{-0.527}]; $p<\num{0.001}$). Moreover, we used a state-of-the-art machine learning model to calculate positive and negative sentiment scores for the fact-checked posts (see \Cref{supp:descriptives}). Here, we found that the efficacy of community notes was robust and had no significant difference between posts with low negative sentiment (ATT: \num{-0.622}; 99\% CI: [\num{-0.629}, \num{-0.615}]; $p<\num{0.001}$) and those with high negative sentiment (ATT: \num{-0.619}; 99\% CI: [\num{-0.625}, \num{-0.612}]; $p<\num{0.001}$), and between posts with low positive sentiment (ATT: \num{-0.619}; 99\% CI: [\num{-0.625}, \num{-0.612}]; $p<\num{0.001}$) and those with high positive sentiment (ATT: \num{-0.622}; 99\% CI: [\num{-0.629}, \num{-0.615}]; $p<\num{0.001}$). Ultimately, we implemented (and validated) a topic modeling approach (see \Cref{supp:extraction}) to study how the efficacy of community notes varies across different topics, namely, $\var{Economy}$, $\var{Health}$, $\var{Politics}$, and $\var{Science}$. Posts that did not fall into one of these topic categories were categorized as \textsc{Other}. Here, we found that the ATT was lower for posts related to economy (ATT: \num{-0.580}; 99\% CI: [\num{-0.596}, \num{-0.564}]; $p<\num{0.001}$), health (ATT: \num{-0.560}; 99\% CI: [\num{-0.576}, \num{-0.544}]; $p<\num{0.001}$), and politics (ATT: \num{-0.578}; 99\% CI: [\num{-0.588}, \num{-0.567}]; $p<\num{0.001}$), as compared to posts not related to economy (ATT: \num{-0.625}; 99\% CI: [\num{-0.630}, \num{-0.620}]; $p<\num{0.001}$), health (ATT: \num{-0.626}; 99\% CI: [\num{-0.631}, \num{-0.621}]; $p<\num{0.001}$), and politics (ATT: \num{-0.636}; 99\% CI: [\num{-0.642}, \num{-0.631}]; $p<\num{0.001}$). This suggests that users have a higher resistance towards community notes if they are attached to posts covering economy, health, and political topics. We found no such statistically significant difference for science-related posts (ATT: \num{-0.613}; 99\% CI: [\num{-0.627}, \num{-0.599}]; $p<\num{0.001}$), as compared to posts not covering science-related topics (ATT: \num{-0.622}; 99\% CI: [\num{-0.627}, \num{-0.617}]; $p<\num{0.001}$).

\begin{figure}
	\captionsetup[subfloat]{font={bf, small}, skip=0pt, singlelinecheck=false, labelformat=simple, position=top}
	\centering
    \subfloat[]{\includegraphics[width = .32\textwidth]{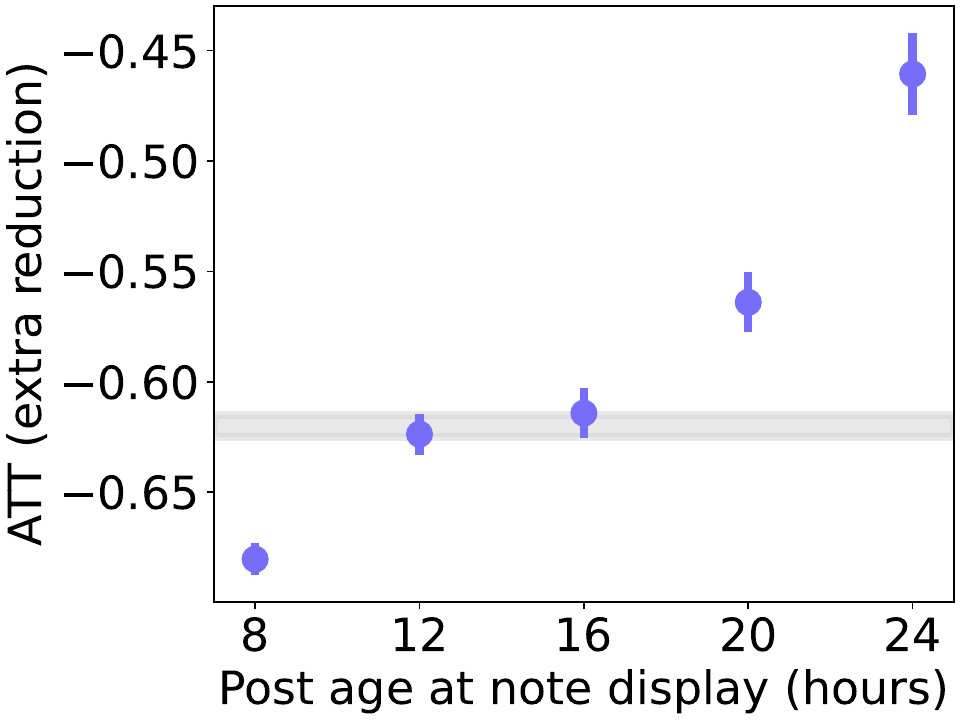}}
    \hspace{.5em}
    \subfloat[]{\includegraphics[width = .32\textwidth]{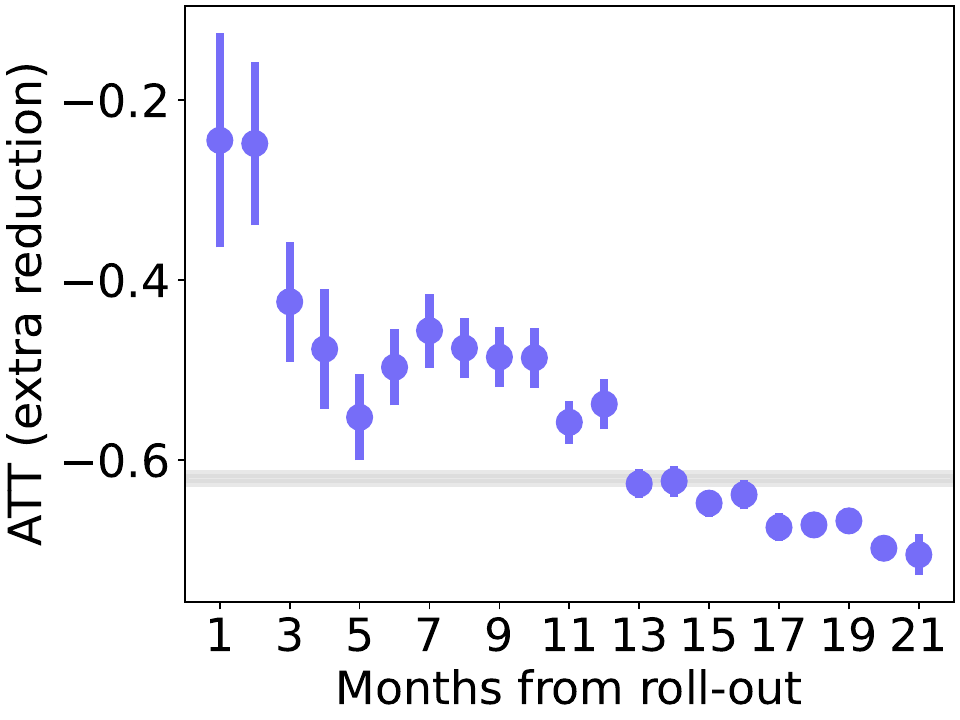}}
    \hspace{.5em}
    \subfloat[]{\includegraphics[width = .32\textwidth]{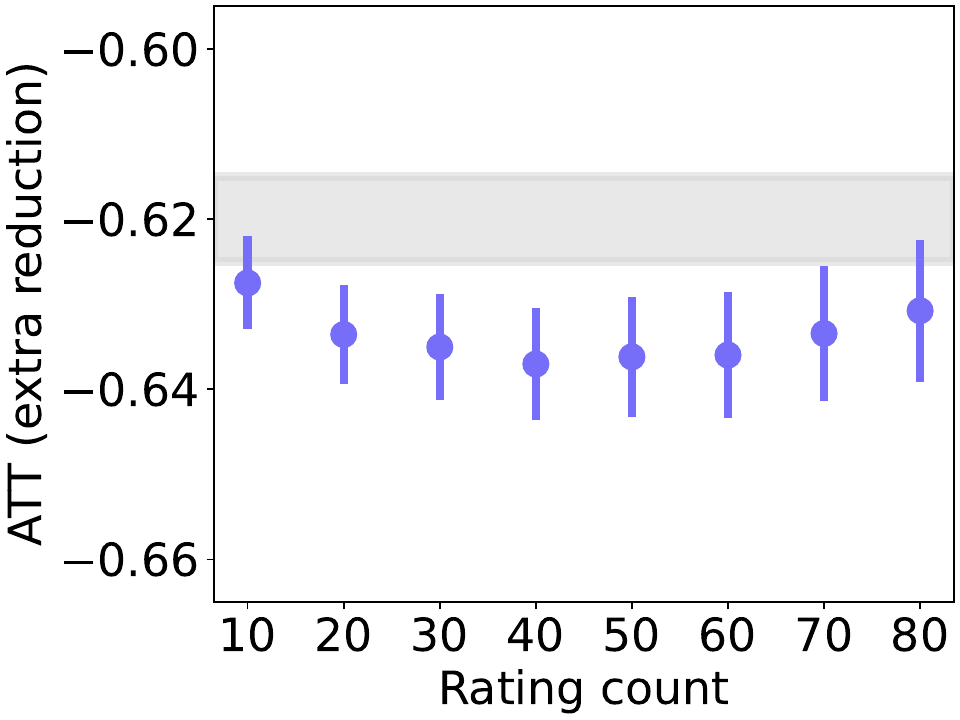}}
    \\
    \subfloat[]{\includegraphics[width = \textwidth]{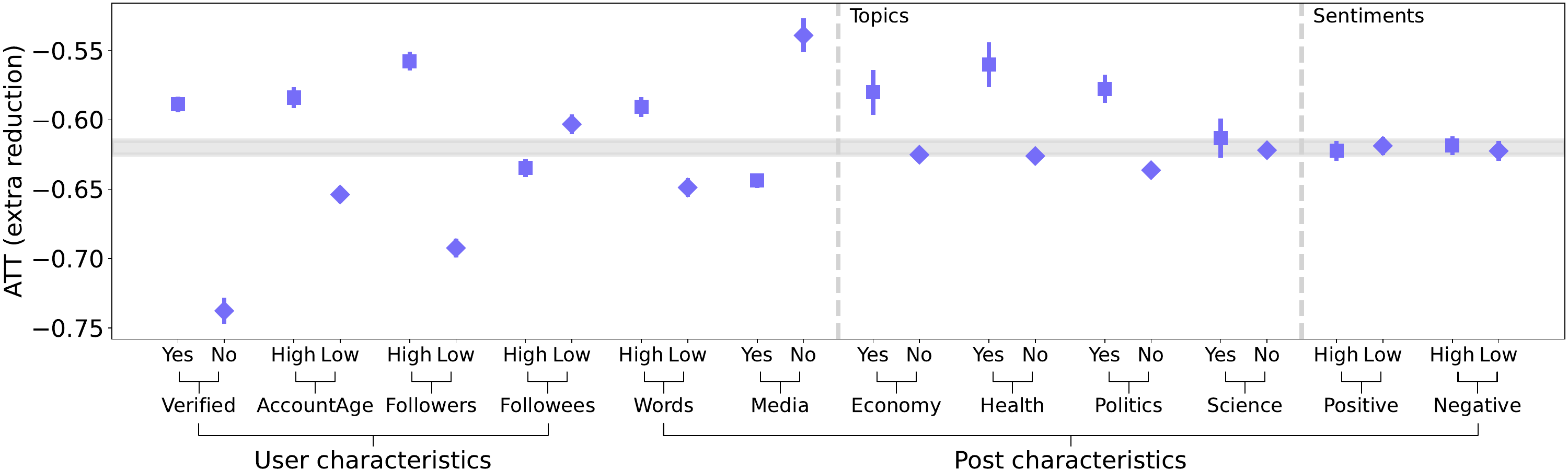}}
	\caption{\textbf{Sensitivity analysis.} \textbf{(a)}~The estimated ATTs across different response times from post creation to note display (grouped within 4-hour windows). Full regression results are reported in \Cref{tab:did_sensitivity_cohorts}. \textbf{(b)}~The estimated ATTs of community notes across the months following the roll-out of ``Community Notes'' program in October 2022. Full regression results are reported in \Cref{tab:did_sensitivity_mfro}. \textbf{(c)}~The estimated ATTs of community notes depending on the number of ratings (10--80) by other fact-checking contributors. Here, we consider only community notes for misleading posts that have never been rated as not helpful. Full regression results are reported in \Cref{tab:did_sensitivity_ratings}. \textbf{(d)}~The estimated ATTs within subgroups separated by user and post characteristics. Full regression results are reported in \Cref{tab:did_sensitivity_user_charcs}, \Cref{tab:did_sensitivity_post_charcs1}, and \Cref{tab:did_sensitivity_post_charcs2}. In all plots, the error bars represent 99\% CIs and the grey bands visualize the ATT (with 99\% CIs) estimated via the two-period DiD model from our main analysis. }
	\label{fig:sensitivity}
\end{figure}

\newpage
\subsection*{Effect on the Cumulative Repost Count}


The results of our DiD analysis imply that displaying community notes can reduce the spread of misleading posts on X by, on average, 62.0\% compared to the control group and relative to the before-display period. However, we also found that the half-life of reposts over 36 hours is 5.75 hours with only 13.5\% of all helpful notes displayed before this time point (\Cref{fig:data_overview}d). This indicates that community notes, on average, might come too late to reduce the spread of misleading posts at the most viral stages of diffusion. To further scrutinize the actual real-world effect of community notes, we also analyzed the share of the cumulative repost count for misleading posts actually prevented by the community notes on X. 


To this end, we used our DiD models to predict the number of reposts (at 15-minute intervals) that the treated posts would have received in the absence of displayed notes. Subsequently, we compared the ratio of the predicted cumulative repost count (over 36 hours) in the absence of displayed notes to the actual cumulative number of reposts (over 36 hours). \Cref{fig:simulation}a shows the actual vs. predicted ratio across different response times from post creation to note display. As expected, we found that the ratio of reposts prevented due to community notes decreased over the response time, \ie, was lower for late vs. early notes. The average reduction in the actual vs. predicted ratio is \num{-0.153}, which means that community notes -- in its current implementation -- reduced the overall number of reposts on X by merely 15.3\%. Similarly, in absolute numbers, we only observed a relatively small reduction of, on average, \num{446} reposts between the actual cumulative repost count (mean of \num{1793})
compared to the predicted cumulative repost count (mean of \num{2239}; $\var{KS} = \num{0.040}, p < 0.001$; see \Cref{fig:simulation}b/c).

Additionally, we simulated the potential reduction in the cumulative repost count if all community notes \emph{would} have been displayed at specific times from 2 to 36 hours after post creation. As shown in \Cref{fig:simulation}d, if community notes would have been displayed on misleading posts at the second hour since post creation, they would have reduced the overall repost count by 53.1\% ($p<\num{0.001}$). The reduction gradually decreased with increasing post ages at note display and became statistically insignificant if more than 24 hours have passed since the creation of the posts.

\begin{figure}
	\captionsetup[subfloat]{font={bf, small}, skip=0pt, singlelinecheck=false, labelformat=simple, position=top}
	\centering
    \subfloat[]{\includegraphics[width = \textwidth]{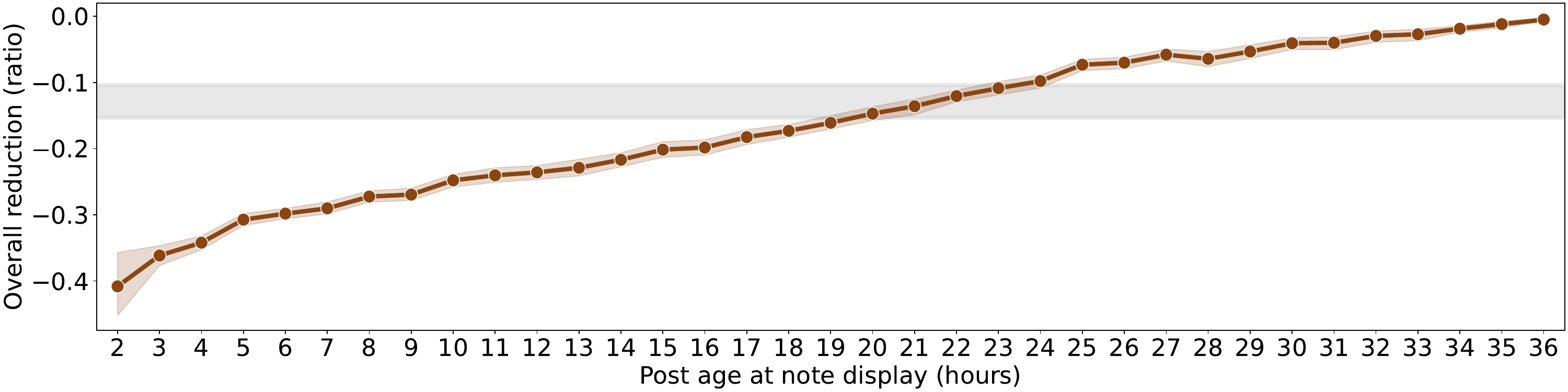}}
    
	\subfloat[]{\includegraphics[width = .32\textwidth]{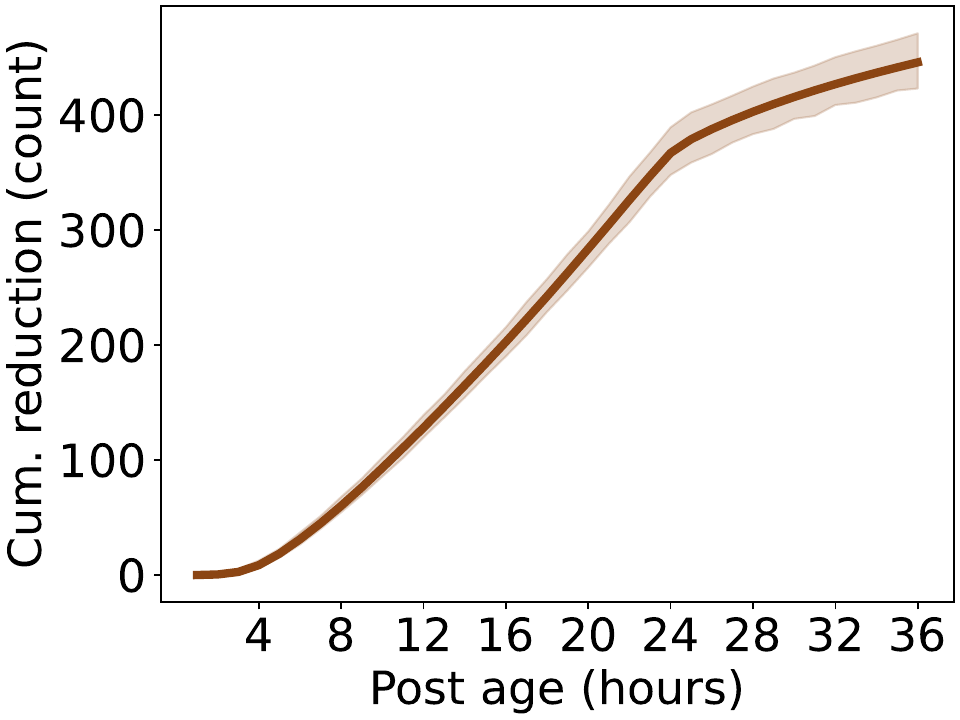}}
	\hspace{.5em}
    \subfloat[]{\includegraphics[width = .32\textwidth]{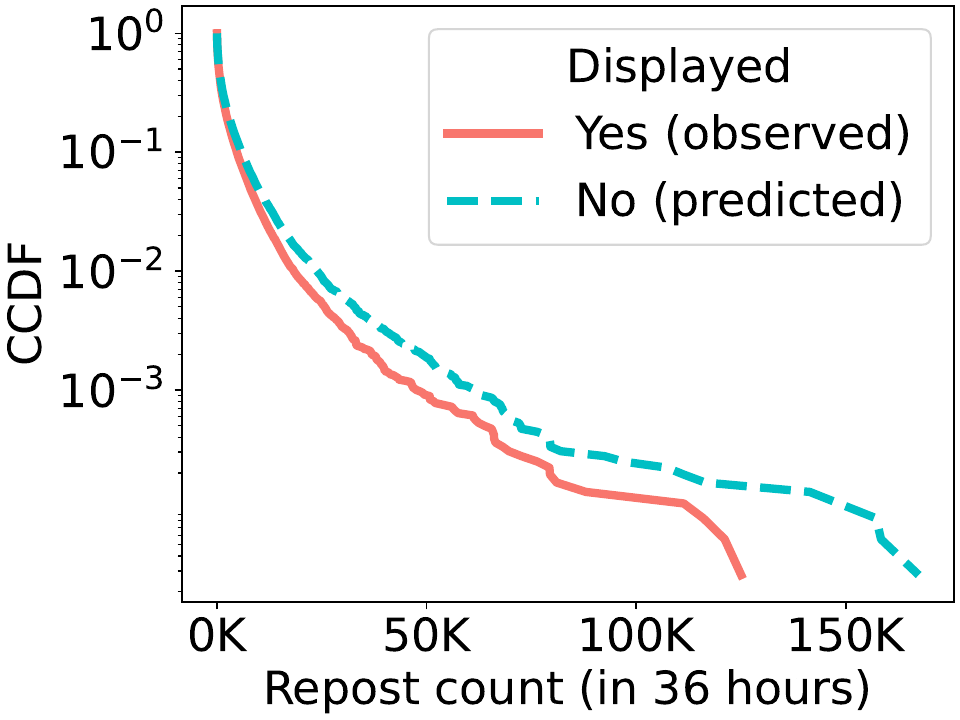}}
	\hspace{.5em}
	\subfloat[]{\includegraphics[width = .32\textwidth]{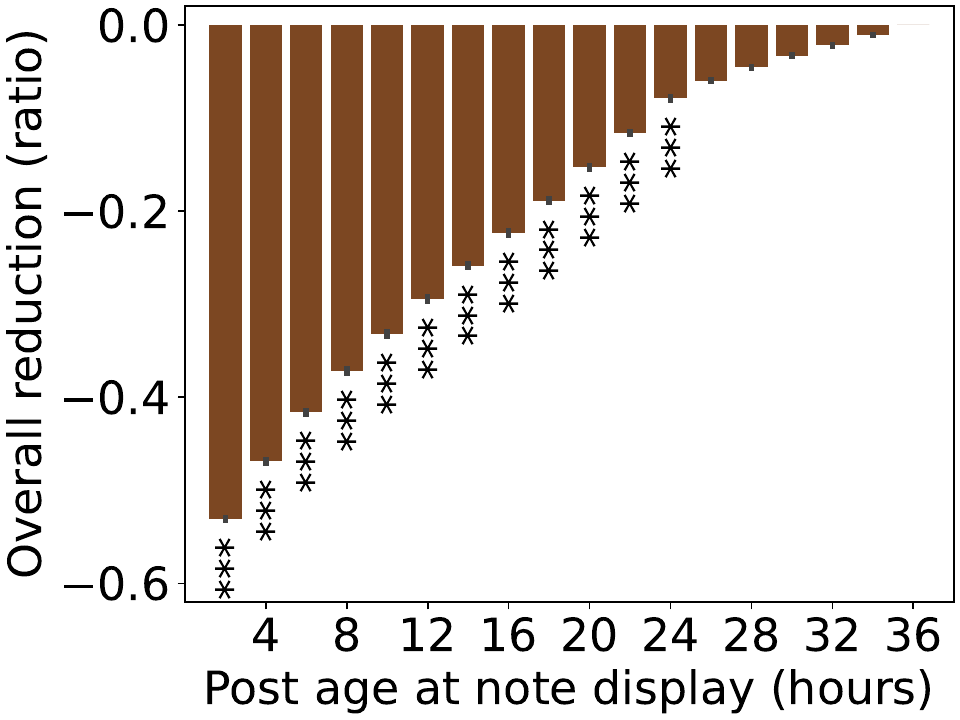}}
	\caption{\textbf{Effect of community notes on the cumulative repost count of misleading posts on X.} \textbf{(a)}~The changes in the ratio of the reduction of reposts relative to the predicted overall repost count over the response time from post creation to note display (grouped within 1-hour windows). The grey band ranges between the mean and the median of the ratio of the reduction. \textbf{(b)}~The estimated cumulative count of reposts that community notes prevents at different post ages. The error band represents 99\% CIs. \textbf{(c)}~CCDFs showing the actually observed repost count for source posts with displayed community notes and the predicted repost count that the source posts would have received in the absence of community notes display. \textbf{(d)}~The estimated overall reduction in reposts if all posts with community notes would have been displayed simultaneously from 2 to 36 hours after post creation. Statistical significance ($^{*}p<0.01$; $^{**}p<0.005$; $^{***}p<0.001$) was calculated using two-tailed KS tests.}
	\label{fig:simulation}
\end{figure}

\newpage
\subsection*{Effect on Post Deletion}

We further examined the treatment effect of community notes on the deletion of misleading posts. Notably, the X API only provided us with the deletion status and did not offer access to information regarding the time of deletion. Therefore, we did not know whether the posts were deleted before or after the display of community notes. Given this, we employed Regression Discontinuity Design (RDD) and hypothesized that, in the absence of a significant effect of community notes, the probability of post deletion should have exhibited smooth continuity (or keep stable) around the threshold of the note helpfulness score that determined the display of community notes (see \Cref{supp:deletedPosts} for details). As of the date of data collection, the cut-off point (threshold) for the note helpfulness score was 0.4, meaning that only community notes receiving a score of 0.4 or above are eligible to be displayed on the corresponding misleading posts. Thus, in our RDD model, observations are assigned to the treatment or control group based on their values of the note helpfulness score, and the treatment effect is estimated by comparing the outcomes just above and just below the cutoff point.

The estimated changes in the ratio of deleted posts for different note helpfulness scores are visualized in \Cref{fig:note_score_deletion}. We observed a clear discontinuity at the cut-off point of 0.4. Specifically, the estimated treatment effect of community notes display in our RDD model was \num{1.034} (99\% CI: [\num{0.800}, \num{1.297}]; $p<0.001$), which implies that the odds for deletion were 103.4\% higher for posts with displayed community notes than for posts without displayed community notes. Additionally, the effect of community notes on post deletion was higher for high-quality fact-checks, as indicated by higher deletion ratio for notes with higher helpfulness scores (see \Cref{supp:deletedPosts} for details and full estimation results)

\begin{figure}
	\captionsetup[subfloat]{font={bf, small}, skip=0pt, singlelinecheck=false, labelformat=simple, position=top}
	\centering
    \includegraphics[width = \textwidth]{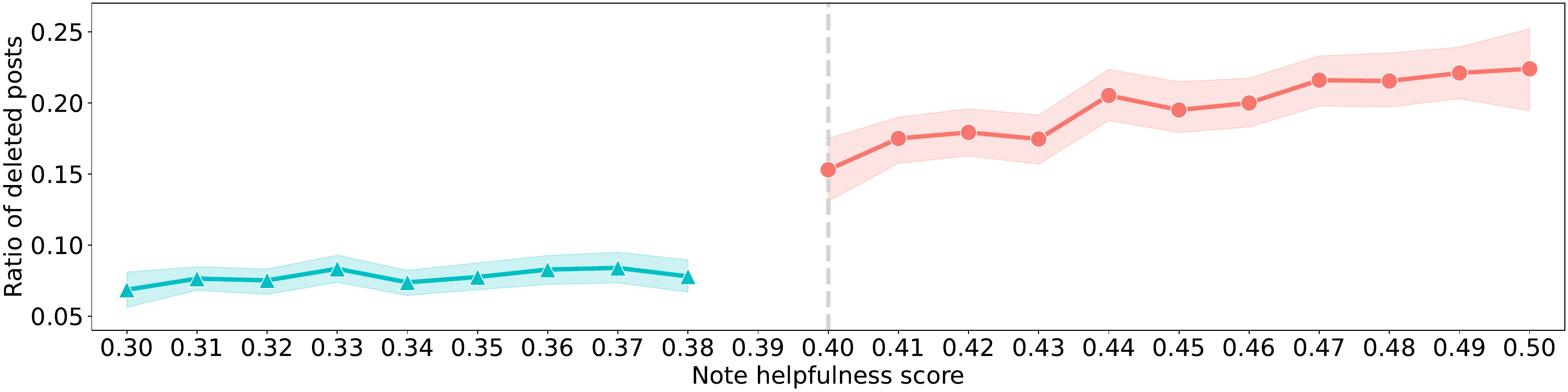}
	\caption{\textbf{Effect of community notes on the deletion of misleading posts on X.} Shown are the ratios of deleted posts for different values of note helpfulness. Note helpfulness scores are re-centered based on the cutoff point of 0.4. Only community notes with note helpfulness scores of 0.4 or above are displayed on the corresponding misleading posts. The note helpfulness scores are rounded to two decimal places. Notes with helpfulness scores of 0.39 are omitted to prevent treatment contamination due to fluctuations between recalculated note helpfulness scores and note helpfulness scores used in production. The error bands represent 99\% CIs. See \Cref{supp:deletedPosts} for further details and full estimation results.}
	\label{fig:note_score_deletion}
\end{figure}

\clearpage
\section*{Discussion}
\label{sec:discussion}

We performed a large-scale quasi-experimental study to assess the real-world efficacy of community-based fact-checking on the social media platform X. 
Based on time series data for $N=\,${\num{237677}} (community fact-checked) X cascades that had been reposted more than \num{431} million times, we found that exposing users to community notes reduced the spread of misleading posts by, on average, 62.0\%. Furthermore, the odds for post deletion were 103.4\% higher for posts with displayed community notes than for posts without displayed community notes. While the treatment effect of community notes was pronounced and statistically significant across the board, we still observed substantial variations concerning both user characteristics and content characteristics. Specifically, the efficacy of community-based fact-checking was discounted for posts originating from influential accounts (\eg, verified users, high-follower accounts). A potential explanation is that such users may have the ability to attract more like-minded individuals, fostering echo chambers where their false beliefs are reinforced  \cite{mosleh2022measuring}, thereby potentially mitigating the efficacy of community notes. Content characteristics also significantly influenced the efficacy of community notes. For example, the presence of attached media items, such as images or videos, significantly amplified the intervention effect. There were also significant differences across topics: community notes showed a lower efficacy for economy-related, health-related, and political misinformation. Altogether, our findings complement earlier studies focusing on the question of whether crowds are able to accurately assess social media content \cite{Micallef.2020,Bhuiyan.2020,Pennycook.2019,Epstein.2020,Allen.2020,Allen.2021,Godel.2021,Martel.2023}, by offering real-world empirical evidence that community notes reduce the spread of misleading posts on X.


While community notes exhibited a pronounced treatment effect, the timing of their display poses a significant challenge. Compared to the rapid dissemination of posts on X, fact-checking via community notes was relatively slow. In our data, the half-life of reposts over 36 hours is 5.75 hours with only 13.5\% of all helpful notes displayed before this time point. The long delay in displaying community notes rendered them ineffective in curtailing the spread of misinformation during the early, and often most viral, stages of diffusion. Specifically, our study shows that community notes -- in its current implementation -- reduced the overall number of reposts on X by merely 15.3\%. The crucial implication here is that community-based fact-checking interventions need to be faster in their response to misinformation dissemination to make a more substantial impact. 


As with any study, ours is not without limitations, which present opportunities for future research. Firstly, our focus on community-based fact-checking was limited to one social media platform, specifically examining the effectiveness of community notes on X. While this platform represents the only large-scale implementation of community-based fact-checking on mainstream social media platforms to date, future research could expand to include other platforms if they choose to implement similar features (\eg, YouTube has recently begun testing a fact-checking feature comparable to X's Community Notes). Additionally, the Community Notes feature itself may evolve over time, potentially influencing its effectiveness in combating misinformation. Consequently, findings from our study may need to be revalidated in the future to account for any developments or enhancements to this feature. Moreover, our study primarily focused on the American context and the global North, with a specific emphasis on posts in English. Future research could explore the efficacy of community-based fact-checking across different cultural contexts to gain a more comprehensive understanding of its impact on diverse communities. Lastly, it could be valuable to investigate how the intervention effect varies across different demographic groups, such as age, gender, political orientation, or education level, to uncover insights into who is most influenced by community-based fact-checking efforts.


Fact-checking becomes increasingly vital, particularly in light of emerging challenges posed by the scalability of AI-generated misinformation. Countermeasures against misinformation are more relevant than ever, and community-based interventions, such as community notes, hold promise in this fight, at least in concept. However, the effectiveness of such interventions hinges on their ability to respond swiftly to the dynamic nature of misinformation dissemination. Therefore, enhancing the speed and efficacy of community-based fact-checking approaches is crucial in fortifying our defenses against the spread of false information in an era marked by heightened digital influence.


\clearpage
\section*{Methods}
\label{sec:methods}

\subsection*{Data Collection}
\label{sec:data} 

We downloaded all community note contributions via a dedicated website on June 16, 2024 \cite{Twitter.2024a}. X routinely updates the community note contributions on a daily base and releases them to the public via this website. The community note contributions were documented in three distinct datasets: notes, ratings, and note status history. Each note in the notes dataset includes its unique note ID, referenced post ID, creation timestamp, and classification (\ie, ``misinformed or potentially misleading'' or ``not misleading''). The ratings dataset includes all the ratings submitted by the contributors along with their creation timestamps. The note status history dataset records the changes in the status of referenced notes and corresponding timestamps. The recorded note statuses contain the information on the helpfulness ratings, \ie, whether the notes were displayed on the corresponding misleading posts.

Based on the referenced post IDs in the notes dataset, we collected the corresponding source posts via the X API v2 post lookup endpoint. Here, we only considered source posts written in English. Each post object returned from this endpoint contains detailed information related to the post itself (such as text, creation time, and media metadata) and its author profile (such as verified status, number of followers, and number of followees). Subsequently, we employed the X API v2 full-archive post counts endpoint to collect \emph{time series data} of repost counts at minute level for the community fact-checked posts over a period of 36 hours since their creation. In total, our dataset contains {time series data} of repost counts for a total of $N =$ \num{237677} (community fact-checked) posts that were created between October 6, 2022 and June 11, 2024 and had been reshared more than \num{431} million times on X (see \Cref{supp:data_summary} for descriptive statistics).  

\subsection*{Empirical Analysis}


We employed a Difference-in-Differences design to estimate the treatment effect of community notes on the spread of misleading posts on X as measured by the repost count over time following the note display. The treatment group consists of all source posts with displayed notes, \ie, posts that received a community note rated as helpful and, thus, displayed to all users on X. We considered the time of the display of the first helpful note for each post as the start of the treatment. 

To ensure comparability, we constructed a control group from the source posts that share similar characteristics with the source posts in the treatment group but have no displayed notes. Given that the number of source posts without displayed notes (\num{201604}) is nearly 6 times the number of source posts with displayed notes (\num{36073}), we performed one-to-one propensity score matching for source posts with displayed notes and find their closest matches in the source post without displayed notes to construct a balanced control group (see \Cref{supp:group_construction}). The one-to-one matching was conducted based on the variables from user profiles (\eg, followers and followees) and post features (\eg, sentiments and topics). As a result, the source posts in the control group have no statistically significant difference from the source posts in the treatment group with respect to the user and post characteristics. Subsequently, we assigned the virtual time of note display to the source posts in the control group according to the corresponding posts in the treatment group and recentered the repost timelines around the display of community notes. 


\textbf{Two-period DiD:} We implemented a two-period DiD model to estimate differences in repost counts between the treatment and control group before and after the display of community notes. In our data, community notes were displayed \num{1.239} hours after post creation at the earliest, and only 5.2\% of community notes were displayed within 4 hours since post creation. Therefore, we considered a 4-hour window before the note display as before-display period. The after-display period was set to between 1 and 12 hours from the note display. Notably, the time points when the community notes were displayed were not exactly corresponding to the time points of the collected repost time series. Therefore, we omitted one hour preceding the display of community notes to prevent potential contamination of community notes display on before-display period. The dependent variable in the DiD model is the number of reposts (\ie, $\var{RepostCount}$). Given that $\var{RepostCount}$ is a non-negative count variable with overdispersion, we employed a negative binomial regression. Formally, we specified the negative binomial regression model with post-specific random effects (\ie, random intercepts) for the two-period ATT estimation as:
\begin{equation}
\begin{aligned}
    log(E(RepostCount_{it}|\bm{x_{it}})) = & \beta_{0} + \var{\beta_{1}Display_{i}} + \var{\beta_{2}After_{t}} + \var{\beta_{3}Display_{i} \times After_{t}} + \\
    & \var{\beta_{4}PostAge_{it}} + \var{\mu_{post}}.
\end{aligned}
\end{equation}
where $\bm{\var{x_{it}}}$ indicates all the independent variables and $\var{\beta_{0}}$ is the intercept. $\var{Display_{i}}$ is a group-specific dummy variable indicating whether the source post $\var{i}$ belongs to the treatment group ($=$ 1) or control group ($=$ 0). $\var{After_{t}}$ is a time-specific dummy variable indicating the time is before ($=0$) or after the note display ($=1$). $\var{Display_{i} \times After_{t}}$ is the difference-in-differences interaction capturing the treatment effect. The variable $\var{\mu_{post}}$ represents post-specific random effects, which capture post-level heterogeneity that cannot be reflected by the general post features in the propensity score matching. For example, posts related to real-world events that happened suddenly may get more engagement.

\textbf{Multi-period DiD:} We estimated a multi-period DiD to enable a temporal analysis of the treatment effect. Here, we again considered a 4-hour window before the note display as the before-display period and then examined the hourly multi-period ATTs from 1 to 12 hours after the note display. We again excluded the first hour before the note display to prevent potential treatment contamination. The baseline of time effects is thus the second hour before note display. Formally, we specified the negative binomial regression model for the multi-period ATT estimation as:
\begin{equation}
\begin{aligned}
    log(E(RepostCount_{it}|\bm{x_{it}})) = & \var{\beta_{0}} + \var{\beta_{1}Display_{i}} + \bm{\var{\beta_{2}^{'}Before_{t}}} + \bm{\var{\beta_{3}^{'}After_{t}}} + \\
    &\var{\bm{\beta_{4}^{'}}Display_{i}} \times \bm{\var{Before_{t}}} + \var{\bm{\beta_{5}^{'}}Display_{i}} \times \bm{\var{After_{t}}} +\\
    &\var{\beta_{6}PostAge_{it}} + \mu_{post},
\end{aligned}
\end{equation}
where $\bm{\var{\beta_{4}}}$ is the estimates for the parallel test, and $\bm{\var{\beta_{5}}}$ is the estimates for the corresponding difference-in-difference interactions that can be transformed to multi-period ATTs. Further, $\var{PostAge_{it}}$ denotes the age of source post $\var{i}$ at time $\var{t}$ since its creation. All other variables are the same as those in the two-period model.

\textbf{Calculation of ATTs:}
The coefficients estimated for the difference-in-differences terms in the two-period and multi-period regression models are the natural logarithms of the ratios for the number of reposts with the treatment of note display compared to the number of reposts that are expected to receive without the note display during the after-display period. We exponentially transformed the coefficient estimates of the difference-in-differences terms in the models and measured the treatment effect of community notes (\ie, ATT) as:
\begin{equation}
\begin{aligned}
    ATT = e^{\beta} - 1,
\end{aligned}
\end{equation}
where $\beta$ is the coefficient estimates for the difference-in-differences terms. $\var{ATT}$ indicates the ratio of extra change of the reposts in the source posts with displayed notes relative to the reposts that the source posts are expected to receive without displayed notes. We used this indicator to analyze the efficacy of community notes.

\textbf{Implementation:}
We used Python 3.8.6 to conduct our empirical analyses on the High-Performance Computing platform at the University of Luxembourg. Our regression models were implemented using the \emph{pystata} Python package with Stata 17.0 MP-parallel edition (2-core network). 

\subsection*{Ethics Statement}

This research has received ethical approval from the Ethics Review Panel of the University of Luxembourg (ref. ERP 23-053 REMEDIS). All analyses are based on publicly available data.

\clearpage
\section*{Additional information}

\vspace{0.4cm}
\noindent
\textbf{Code availability.} Upon publication of this work, the analysis scripts to reproduce our study will be made available via OSF.

\vspace{0.4cm}
\noindent
\textbf{Data availability.} Upon publication of this work, all data and materials needed to recreate the analysis will be made available via OSF.

\section*{Author information}

\textbf{Author contributions.} Y.C., M.P., T.R., D.R., A.T., G.L., and N.P. conceived and designed the experiments. M.P. and T.R. collected the data. Y.C., M.P., and T.R. analyzed the data. Y.C., M.P., T.R., D.R., A.T., G.L., and N.P. wrote the manuscript. All authors approved the manuscript.

\vspace{0.4cm}
\noindent
\textbf{Correspondence.} Nicolas Pröllochs (\url{nicolas.proellochs@wi.jlug.de})

\vspace{0.4cm}
\noindent
\textbf{Competing interests.} The authors declare no competing interests.

\section*{Acknowledgments}

This study was supported by research grants from the German Research Foundation (DFG grant 492310022), the Luxembourg National Research Fund (INTER\_FNRS\_21\_16554939\_REMEDIS), and the HEC Paris Foundation. The funders had no role in study design, data collection and analysis, decision to publish or preparation of the manuscript.


\clearpage
\bibliography{literature.bib}


\clearpage
\appendix

\renewcommand{\thetable}{S\arabic{table}}
\renewcommand{\thefigure}{S\arabic{figure}}
\setcounter{figure}{0}
\setcounter{table}{0}

\begin{minipage}[t]{\textwidth}
	\nolinenumbers
	\begin{center}
		\huge\bfseries Supplementary Materials
	\end{center}
	\vspace{1cm}
\end{minipage}

\tableofcontents

\clearpage

\section{Descriptive Statistics}
\label{supp:descriptives}

\subsection{Data Overview}
\label{supp:data_summary}

An overview of our dataset is reported in \Cref{tab:data_summary}. In total, the data consists of \num{237677} English source posts that were fact-checked between October 6, 2022 and June 11, 2024, \ie, during an observation period of approximately 20 months. These source posts were authored by \num{60928} unique users and reposted more than \num{431} million times at the time of data collection.

We collected several variables to control for user characteristics that could potentially affect engagement with the posts. These variables include $\bm{\var{Verified}}$, which is a binary indicator showing whether the author is verified ($=$ \num{1}) or not ($=$ \num{0}). Additionally, $\bm{\var{AccountAge}}$ represents the age of the post account in days at the time of data collection. Furthermore, $\bm{\var{Followers}}$ denotes the number of followers for the post author, while $\bm{\var{Followees}}$ indicates the number of followees for the post author.

Given that the engagement with posts might change following the development of ``Community Notes'' program, we controlled for the time length from the months in which the posts were created to the roll-out of the program, \ie, Months From the Roll-Out ($\bm{\var{MFRO}}$). Additionally, we extracted a wide range of content characteristics that may affect engagement with the source posts. These characteristics include $\bm{\var{Media}}$, a binary variable indicating whether the post contains media elements ($=$ \num{1}) or not ($=$ \num{0}). Moreover, $\bm{\var{Words}}$ represents the word count in the source post. To determine the number of words in the source post, we performed text preprocessing by removing user mentions from the beginning of each text and eliminating all URLs. Furthermore, we converted HTML entities to Unicode characters and applied general International Components for Unicode (ICU) transforms \cite{ICU.docs} to further normalize the character set. We then used ICU BreakIterators to identify word boundaries for splitting the texts before counting the number of tokens containing alphanumeric characters. Additionally, we computed sentiment scores for the posts in our dataset using the Twitter-RoBERTa-base model, which has been fine-tuned for sentiment analysis on tweets and achieved a high performance compared to other models  \cite{Loureiro.2022}. We use this pre-trained machine learning model to calculate the probabilities that a given post conveys positive emotion ($\bm{\var{Positive}}$) and negative emotion ($\bm{\var{Negative}}$).

Ultimately, we performed topic modeling to study heterogeneity across topics.  Specifically, we employed (and validated) a supervised machine learning framework (see details in \Cref{supp:extraction}) to categorize the source tweets into predefined topics: (i) $\bm{\var{Politics}}$, (ii) $\bm{\var{Health}}$, (iii) $\bm{\var{Economy}}$, (iv) $\bm{\var{Science}}$. Posts that did do not fall into one of these topic categories were categorized as \textsc{Other}. 

\begin{table}
    \caption{Dataset overview. Descriptive statistics are reported for the whole dataset (column (1)) and the subsets of posts that have received (column (2)) or have not received a displayed helpful note during our observation period (column (3)). Continuous values are reported as means with standard deviations in parentheses, unless otherwise stated. }
    \footnotesize
    \setlength\tabcolsep{7pt}
    \begin{tabularx}{\textwidth}{Xccc}
        \toprule
        \multicolumn{2}{c}{ } & \multicolumn{2}{c}{Posts with displayed notes} \\
        \cmidrule(l{3pt}r{3pt}){3-4}
         & All & Yes & No\\
        \midrule
        \hspace{1em}\#Reposts & \num{431780333} & \num{64682945} & \num{367097388}\\
        \hspace{1em}\#Posts & \num{237677} & \num{36073} & \num{201604}\\
        \hspace{1em}\#Users & \num{60928} & \num{13847} & \num{55346}\\
        \hspace{1em}Post date & {10/06/2022~--~06/11/2024} & {10/06/2022~--~06/10/2024} & {10/06/2022~--~06/11/2024}\\
        \hspace{1em}Note date & {10/06/2022~--~06/13/2024} & {10/06/2022~--~06/13/2024} & {10/06/2022~--~06/13/2024}\\
        \midrule
        \multicolumn{4}{l}{\textit{User characteristics}}\\
        \midrule
        \hspace{1em}$\var{Verified}$ & \SI{75.4}{\percent} & \SI{76.3}{\percent} & \SI{75.3}{\percent}\\
        \hspace{1em}$\var{AccountAge}$ (days) & \num{2956.865} (\num{1888.894}) & \num{2729.643} (\num{1845.162}) & \num{2997.522} (\num{1893.744})\\
        \hspace{1em}$\var{Followers}$ & \num{2322474} (\num{15263927}) & \num{1513771} (\num{10907782}) & \num{2467175} (\num{15913818})\\
        \hspace{1em}$\var{Followees}$ & \num{6861} (\num{34140}) & \num{7274} (\num{29469}) & \num{6787} (\num{34909})\\
        \midrule
        \multicolumn{4}{l}{\textit{Post characteristics}}\\
        \midrule
        \hspace{1em}$\var{Words}$ & \num{26.027} (\num{14.888}) & \num{23.365} (\num{14.581}) & \num{26.504} (\num{14.892})\\
        \hspace{1em}$\var{Media}$ & \SI{63.6}{\percent} & \SI{72.0}{\percent} & \SI{62.1}{\percent}\\
        \hspace{1em}Sentiment: $\var{Positive}$ & \num{0.155} (\num{0.267}) & \num{0.167} (\num{0.275}) & \num{0.152} (\num{0.265})\\
        \hspace{1em}Sentiment: $\var{Negative}$ & \num{0.429} (\num{0.341}) & \num{0.396} (\num{0.338}) & \num{0.434} (\num{0.342})\\
        \hspace{1em}Topic: $\var{Economy}$ & \SI{13.0}{\percent} & \SI{12.3}{\percent} & \SI{13.1}{\percent}\\
        \hspace{1em}Topic: $\var{Health}$ & \SI{10.8}{\percent} & \SI{9.7}{\percent} & \SI{11.0}{\percent}\\
        \hspace{1em}Topic: $\var{Politics}$ & \SI{31.9}{\percent} & \SI{25.0}{\percent} & \SI{33.1}{\percent}\\
        \hspace{1em}Topic: $\var{Science}$ & \SI{11.1}{\percent} & \SI{13.8}{\percent} & \SI{10.7}{\percent}\\
        \hspace{1em}Topic: $\var{Other}$ & \SI{45.8}{\percent} & \SI{49.6}{\percent} & \SI{45.1}{\percent}\\
        \hspace{1em}$\var{MFRO}$ & \num{14.239} (\num{4.472}) & \num{13.644} (\num{4.799}) & \num{14.346} (\num{4.402})\\
        \addlinespace[0.3em]
        \bottomrule
    \end{tabularx}
    \label{tab:data_summary}
\end{table}

\clearpage

\subsection{Fact-Checking Activity Over Time}
\label{supp:post_note}

With the development of ``Community Notes'' feature on X, the absolute number of fact-checking notes and fact-checked posts increased significantly (see \Cref{fig:data_overview}b in the main paper). However, the ratio of displayed notes to all community notes over time did not increase and even decreased after January of 2023 (\Cref{fig:crh_note}a). Notably, the ratio of displayed notes was relativly stable around \num{0.1} over the period after April of 2023. On average, 12.0\% community notes were displayed every day, compared to the total notes that were created on the same day (\Cref{fig:crh_note}b).

\begin{figure}
	\captionsetup[subfloat]{font={bf, small}, skip=0pt, singlelinecheck=false, labelformat=simple, position=top}
	\centering
	\subfloat[]{\includegraphics[width = .32\textwidth]{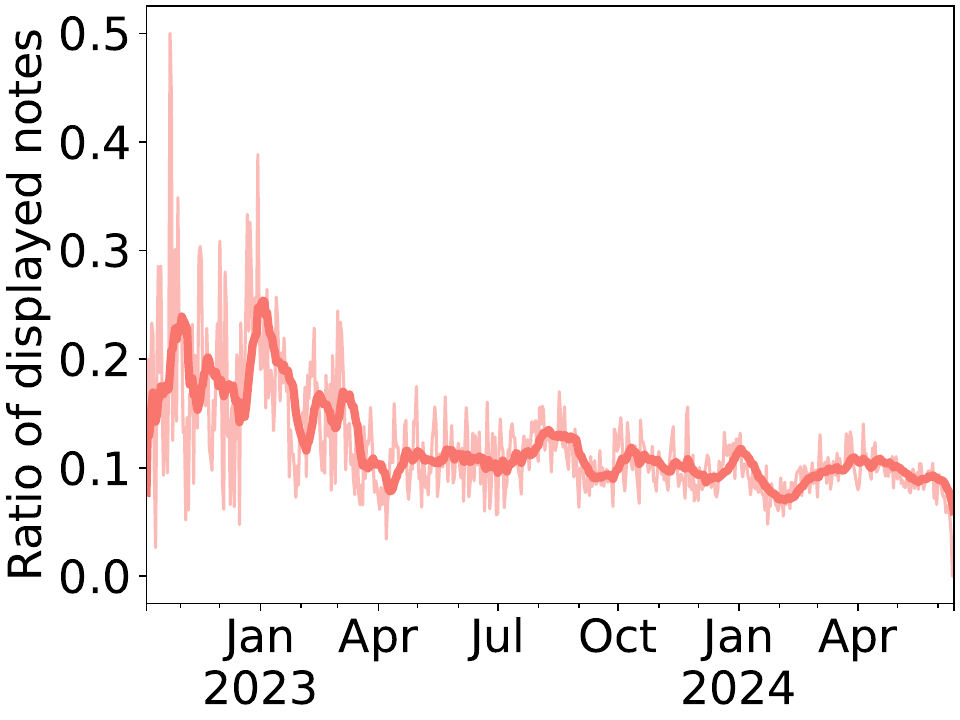}}
	\hspace{.5em}
	\subfloat[]{\includegraphics[width = .32\textwidth]{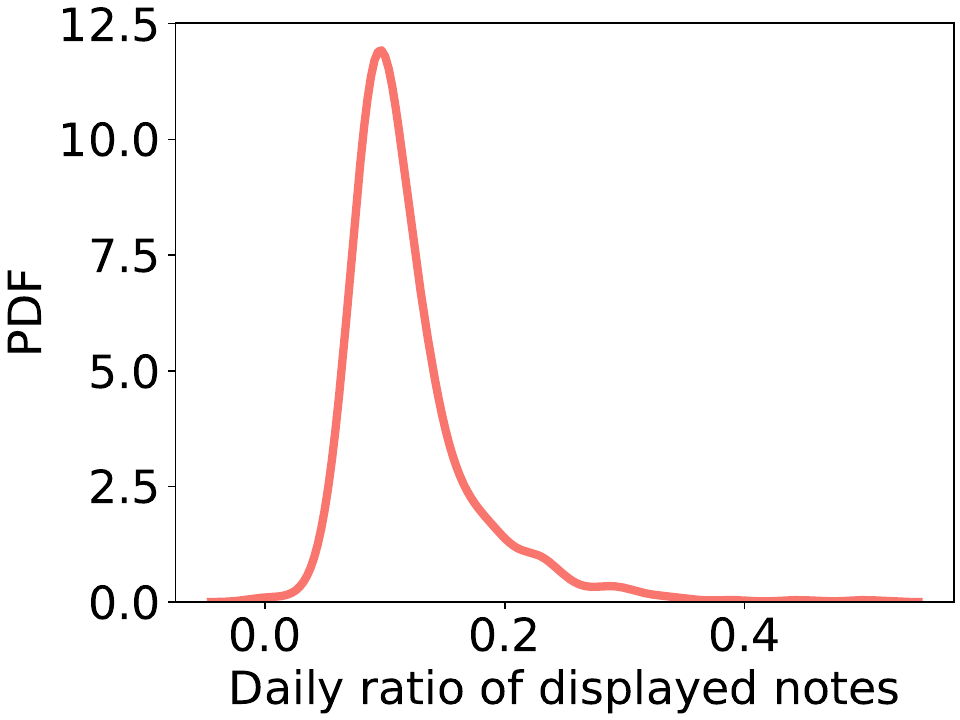}}
	\caption{\textbf{The ratio of displayed notes relative to all the community notes over time.} \textbf{(a)}~Two-week rolling averages of the ratio of displayed notes over time. \textbf{(b)}~The Probability Density Function (PDF) of daily ratio of displayed notes.}
	\label{fig:crh_note}
\end{figure}

\newpage
\subsection{Stability of Note Status}
\label{supp:note_status}

X updates the status of notes on an hourly base according to its note ranking algorithm \cite{Twitter.2024a}. The rating algorithm considers all the ratings made by contributors from different perspectives to calculate the helpfulness score for each community note, determining its status as ``Currently Rated Helpful,'' ``Needs More Ratings,'' or ``Currently Rated Not Helpful.'' Notably, contributors can also write notes that claim the source posts are not misleading. However, such kind of notes will stay at the status of ``Needs More Ratings'' despite of their high helpfulness scores. Community notes that carry the status of ``Currently Rated Helpful'' are directly displayed on the corresponding potentially misleading posts to inform users (see an example in \Cref{fig:data_overview}a). On the other hand, community notes carrying the status of ``Needs More Ratings'' or ``Currently Rated Not Helpful'' are exclusively visible to the Community Notes contributors and, as such, are not expected to exert a substantial impact on the spread of associated source posts. 

Out of the total notes, \num{370234} (83.3\%) remained in ``Needs More Ratings (NMR)'' status without any changes. \num{29401} (6.6\%) notes were initially rated not helpful in their first NonNMR status. Only \num{44664} (10.1\%) notes were initially rated helpful in their first NonNMR status with exact timestamps, and of them, 70.0\% maintained helpful status in their current status and most recent status, and others changed the status into NMR in their current status. No note was rated helpful first and then changed into the status of not helpful in this dataset. Taken together, despite the low proportion of notes rated as helpful, the status of these notes was highly stable over time.

\newpage
\section{Identification of Topics}
\label{supp:extraction}

We used machine learning to assign topic labels to the post in our dataset. Specifically, we employed (and validated) a supervised machine learning framework to categorize the source tweets into predefined topics: (i) \textsc{Politics}, (ii) \textsc{Health}, (iii) \textsc{Economy}, and (iv) \textsc{Science}. These topics have been identified based on a manual assessment of the posts in our dataset and the selection of topics in previous works (\eg, \cite{Vosoughi.2018}). Posts that did do not fall into one of these topic categories were categorized as \textsc{Other}. 

Our supervised machine learning framework proceeded as follows: First, we used a pre-trained large language model as the basis for our classifier, namely, the TwHIN-BERT (large) model~\cite{Zhang.2022}. This model was pre-trained on a corpus of 7 billion posts from X. Its training methodology incorporates not only text-based self-supervision (\eg, masked language model) but also a social objective derived from the social engagements with posts, thereby enhancing the model's understanding of social contexts. Second, we fine-tuned the model to our task using a manually labeled subset of posts from our dataset. Specifically, we tasked two trained research assistants to assign topic labels to a random subset of \num{1500} community fact-checked posts, with each assistant labeling \num{750} distinct posts. To ensure label accuracy, both assistants also labeled an additional \num{175} posts that the other had already labeled.
The inter-rater reliability assessment resulted in a relatively high macro averaged Cohens's $\kappa$ coefficient of \num{0.711} and an overall agreement of \SI{90.8}{\percent}. Finally, we used the labeled posts to train a deep neural network classifier to predict topic labels for all posts in our dataset. All hyperparameters were tuned using 10-fold cross validation. The model was implemented in Python 3.11.3 using the Transformers Python library (version 4.30.2;~\cite{Wolf.2020}). 

The out-of-sample prediction performance (calculated using 10-fold cross-validation) of the classifier on the manually labeled posts is reported in \Cref{tab:topic_model_performance}. The classifier shows a high accuracy of, on average, \num{0.904}; and a macro-averaged $F_1$ score of \num{0.755}.

\begin{table}
\centering
\caption{Out-of-sample performance in topic prediction. Performance metrics were calculated using 10-fold cross-validation.}
\label{tab:topic_model_performance}
    \begin{tabular}{lrr}
        \toprule
          & Accuracy & $F_1$\\
        \midrule
        \textsc{Economy} & 0.916 & 0.733\\
        \textsc{Health} & 0.944 & 0.845\\
        \textsc{Politics} & 0.844 & 0.816\\
        \textsc{Science} & 0.912 & 0.628\\
        {}[Macro-]Average & 0.904 & 0.755\\
        \bottomrule
    \end{tabular}
\end{table}

\newpage
\section{Propensity Score Matching}
\label{supp:group_construction}

To reduce confounding and ensure parallel trends in the treatment and control groups, especially with respect to unmeasured confounders related to real-world events and selection bias, we performed one-to-one propensity score matching \cite{Guo.2014} to construct a control group from the source posts without displayed notes for the source posts with displayed notes. Specifically, we trained a logistic regression model based on the user-level and post-level variables (see \Cref{tab:data_summary}) to estimate the propensity scores of all source posts. All continuous variables were $z$-standardized for better model fitting. According to the propensity score, we matched the closest post in the group without displayed notes for each post in the treatment group without replacement, resulting in the creation of a matched control group for the treatment group. Additionally, we excluded matches from the source post without displayed notes in the first matching and performed the one-to-one propensity score matching again to construct another control group for the treatment group. We considered this second control group as placebo group that also shares similar user and post characteristics with the treatment group. The time when the community notes became displayed in control and placebo groups was assigned according to the corresponding posts in the treatment group. The placebo group can serve as a replacement for the treatment group in the comparison with the control group to demonstrate that there is no additional reduction in reposts without the actual treatment of note display.

As shown in \Cref{fig:psm_rollout}, the distributions of propensity score between source posts with displayed notes and source posts without displayed notes are much more balanced after matching compared to before matching. We tested the performance of the propensity score matching using binary logistic regressions. The binary dependent variable in the logistic regressions is $\var{Display}$ indicating whether the source posts have displayed community notes ($=\num{1}$) or not ($=\num{0}$). The explanatory variables are the user-level and post-level variables from \Cref{tab:data_summary}. Column (1) of \Cref{tab:psm_rollout} reports the regression results for the source posts with and without displayed notes before the matching. Many of the independent variables (\eg, $\var{Verified}$, $\var{AccountAge}$, $\var{Followers}$, and $\var{Media}$) are significantly different between the source posts with and without displayed notes. However, after the matching, all the independent variables among the treatment, control, and placebo groups are well-balanced and have no statistically significant differences (Columns (2) to (4) in \Cref{tab:psm_rollout}).

\begin{figure}
	\captionsetup[subfloat]{font={bf, small}, skip=0pt, singlelinecheck=false, labelformat=simple, position=top}
	\centering
	\subfloat[]{\includegraphics[width = .32\textwidth]{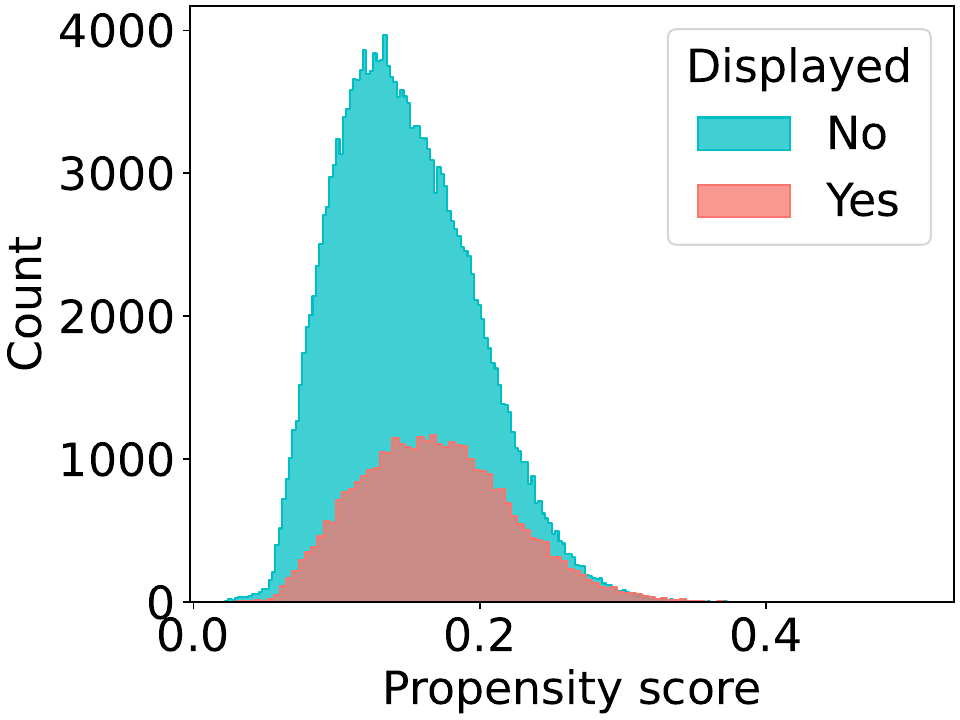}}
	\hspace{.5em}
	\subfloat[]{\includegraphics[width = .32\textwidth]{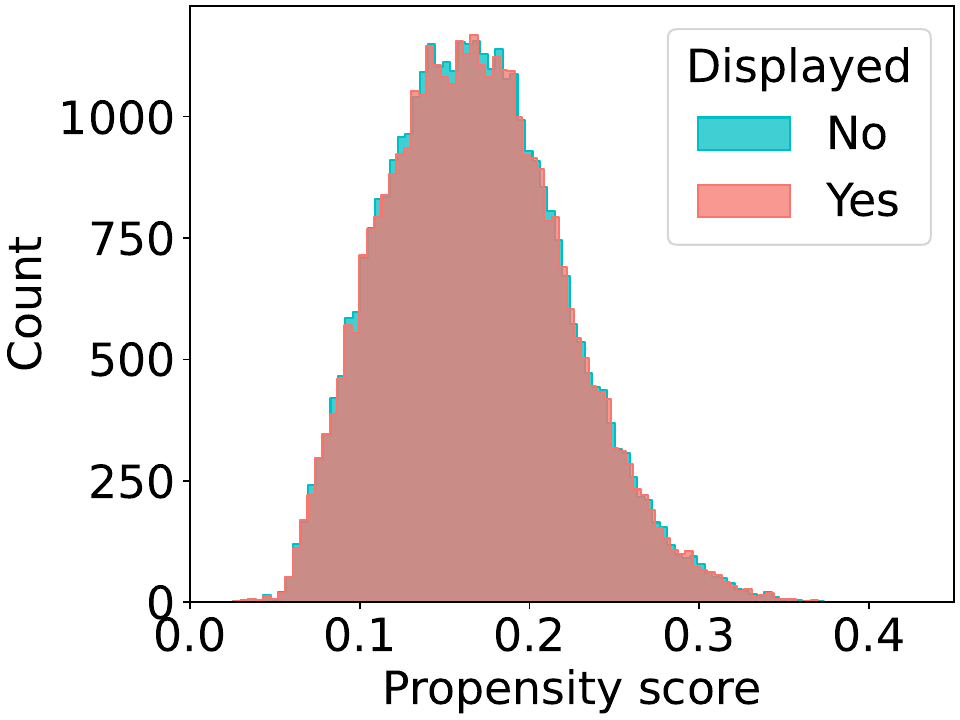}}
    \hspace{.5em}
	\subfloat[]{\includegraphics[width = .32\textwidth]{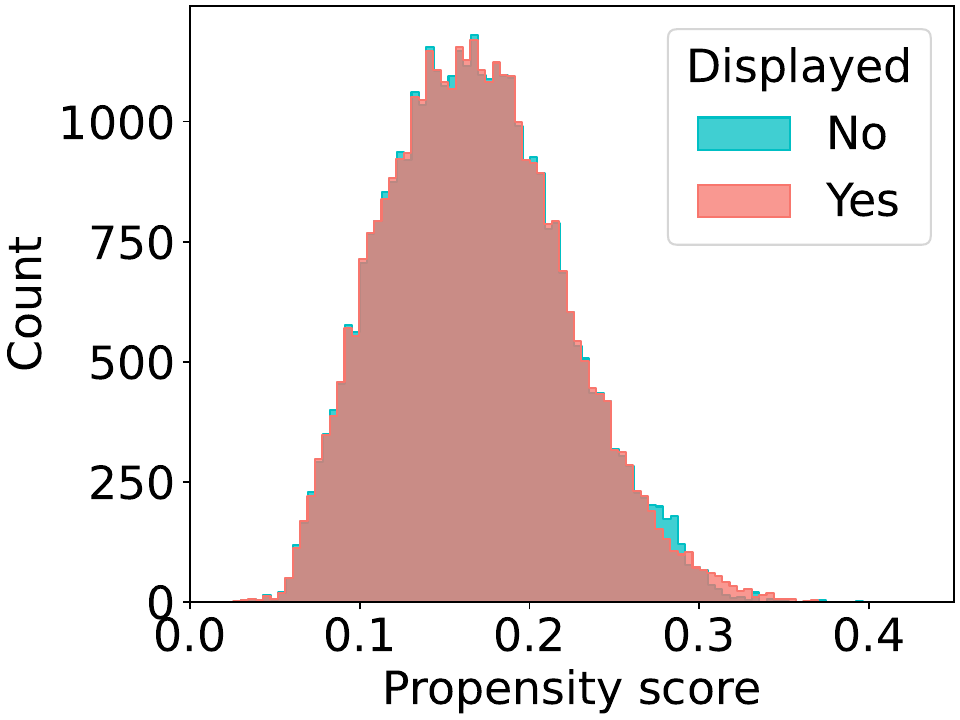}}
	\caption{\textbf{The distributions of propensity score before and after one-to-one propensity score matching.} \textbf{(a)}~The distributions of propensity score in the source posts with displayed notes and source posts without displayed notes before matching. \textbf{(b)}~The distributions of propensity score in the source posts within the treatment group and matched source posts within the control group. \textbf{(c)}~The distributions of peopensity score in the source posts within the treatment group and matched sorce posts within the placebo group.}
	\label{fig:psm_rollout}
\end{figure}

\begin{table}
\centering
\caption{Regression results for the evaluation of propensity score matching. The binary dependent variable is $\var{Display}$. Column (1) reports the results for all the source posts before matching. Column (2) reports the results for the source posts in the treatment and control groups. Column (3) reports the results for the source posts in the treatment and placebo groups. Column (4) reports the results for the source posts in the placebo and control groups. Reported are coefficient estimates with standard errors in parentheses. \sym{*} \(p<0.01\), \sym{**} \(p<0.005\), \sym{***} \(p<0.001\).}
\resizebox{.9\textwidth}{!}{
\begin{tabular}{l*{4}{S}}
\toprule
&\multicolumn{1}{c}{(1)}&\multicolumn{1}{c}{(2)}&\multicolumn{1}{c}{(3)}&\multicolumn{1}{c}{(4)}\\
&\multicolumn{1}{c}{Before matching}&\multicolumn{1}{c}{Treatment : Control}&\multicolumn{1}{c}{Treatment : Placebo}&\multicolumn{1}{c}{Placebo : Control}\\
\midrule
Verified    &       0.127\sym{***}&      -0.003         &      -0.001         &      -0.003         \\
            &     (0.014)         &     (0.018)         &     (0.018)         &     (0.018)         \\
AccountAge  &      -0.085\sym{***}&       0.002         &      -0.007         &       0.009         \\
            &     (0.006)         &     (0.008)         &     (0.008)         &     (0.008)         \\
Followers   &      -0.084\sym{***}&      -0.002         &      -0.017         &       0.016         \\
            &     (0.009)         &     (0.011)         &     (0.011)         &     (0.011)         \\
Followees   &       0.026\sym{***}&       0.009         &       0.003         &       0.005         \\
            &     (0.005)         &     (0.008)         &     (0.008)         &     (0.007)         \\
Words       &      -0.165\sym{***}&       0.005         &      -0.002         &       0.006         \\
            &     (0.006)         &     (0.008)         &     (0.008)         &     (0.008)         \\
Media       &       0.359\sym{***}&       0.027         &       0.001         &       0.027         \\
            &     (0.013)         &     (0.017)         &     (0.017)         &     (0.017)         \\
Economy     &       0.017         &      -0.016         &       0.009         &      -0.024         \\
            &     (0.018)         &     (0.023)         &     (0.023)         &     (0.023)         \\
Health      &      -0.098\sym{***}&       0.008         &      -0.001         &       0.009         \\
            &     (0.020)         &     (0.026)         &     (0.026)         &     (0.026)         \\
Politics    &      -0.262\sym{***}&      -0.006         &       0.006         &      -0.013         \\
            &     (0.014)         &     (0.019)         &     (0.019)         &     (0.019)         \\
Science     &       0.189\sym{***}&      -0.041         &      -0.020         &      -0.021         \\
            &     (0.018)         &     (0.023)         &     (0.023)         &     (0.023)         \\
Positive    &      -0.024\sym{***}&       0.006         &      -0.000         &       0.006         \\
            &     (0.007)         &     (0.009)         &     (0.009)         &     (0.009)         \\
Negative    &      -0.041\sym{***}&       0.007         &      -0.000         &       0.007         \\
            &     (0.007)         &     (0.010)         &     (0.010)         &     (0.010)         \\
MFRO        &      -0.190\sym{***}&       0.001         &       0.001         &       0.000         \\
            &     (0.006)         &     (0.007)         &     (0.007)         &     (0.007)         \\
Intercept      &      -2.028\sym{***}&      -0.007         &      -0.002         &      -0.005         \\
            &     (0.016)         &     (0.021)         &     (0.021)         &     (0.021)         \\
\midrule
\#Posts       &       \num{237672}         &       \num{72136}         &       \num{72136}         &       \num{72136}         \\
\bottomrule
\end{tabular}}
\label{tab:psm_rollout}
\end{table}

\newpage
\section{Estimation Results}
\label{supp:estimation_results}

\subsection{Two-Period ATTs}

The regression results for the two-period ATT estimation are reported in \Cref{tab:did_main_fixed}. The before-display period is between \num{-4} and \num{-2} hours from the note display. The period between 1 and 12 hours from the note display was considered as the after-display period. We omitted one hour preceding the display of community notes to prevent potential contamination of community notes display on before-display period. 

The results for the main regression are reported in Column (1) of \Cref{tab:did_main_fixed}. Furthermore, Column (2) reports the results for an extended regression model that incorporates user and post characteristics as additional explanatory variables. The coefficient estimate for the difference-in-difference term ($\var{Display} \times \var{After}$) 
is qualitatively identical across both models, which further validates the effectiveness of the propensity score matching. 

\begin{table}
\centering
\caption{Regression results for two-period ATT estimation. Post-specific random effects are included. Reported are coefficient estimates with standard errors in parentheses. \sym{*} \(p<0.01\), \sym{**} \(p<0.005\), \sym{***} \(p<0.001\).}
\resizebox{.43\textwidth}{!}{
\begin{tabular}{l*{2}{S}}
\toprule
&\multicolumn{1}{c}{(1)}           &\multicolumn{1}{c}{(2)}           \\
&\multicolumn{1}{c}{Main}&\multicolumn{1}{c}{Extended}\\
\midrule
Display     &       1.286\sym{***}&       1.235\sym{***}\\
            &     (0.023)         &     (0.022)         \\
After   &       0.099\sym{***}&       0.097\sym{***}\\
            &     (0.005)         &     (0.005)         \\
Display$\times$After&      -0.969\sym{***}&      -0.969\sym{***}\\
            &     (0.005)         &     (0.005)         \\
PostAge     &      -0.807\sym{***}&      -0.806\sym{***}\\
            &     (0.003)         &     (0.003)         \\
Verified    &                     &       1.158\sym{***}\\
            &                     &     (0.027)         \\
AccountAge  &                     &      -0.103\sym{***}\\
            &                     &     (0.012)         \\
Followers   &                     &       0.349\sym{***}\\
            &                     &     (0.017)         \\
Followees   &                     &       0.115\sym{***}\\
            &                     &     (0.011)         \\
Words       &                     &       0.076\sym{***}\\
            &                     &     (0.012)         \\
Media       &                     &       0.787\sym{***}\\
            &                     &     (0.026)         \\
Economy     &                     &      -0.300\sym{***}\\
            &                     &     (0.035)         \\
Health      &                     &       0.304\sym{***}\\
            &                     &     (0.038)         \\
Politics    &                     &       0.208\sym{***}\\
            &                     &     (0.027)         \\
Science     &                     &      -0.563\sym{***}\\
            &                     &     (0.033)         \\
Positive    &                     &      -0.089\sym{***}\\
            &                     &     (0.014)         \\
Negative    &                     &       0.197\sym{***}\\
            &                     &     (0.014)         \\
MFRO        &                     &      -0.158\sym{***}\\
            &                     &     (0.011)         \\
Intercept      &       1.203\sym{***}&      -0.181\sym{***}\\
            &     (0.017)         &     (0.033)         \\
\midrule
Post-level RE & \checkmark& \checkmark \\
\midrule
\#Observations       &      \num{614520}   &     \num{614520}\\
\#Posts       &      \num{40968}   &     \num{40968}\\
\bottomrule
\end{tabular}}
\label{tab:did_main_fixed}
\end{table}

\subsection{Parallel Test and Multi-Period ATTs}

The regression results for the parallel test and multi-period ATTs are reported in \Cref{tab:did_main_multi}. We excluded the first hour before the note display to prevent potential treatment contamination. Thus, the second hour before the display of community notes is the baseline period in the multi-period model. The coefficient estimates for the difference-in-differences terms in the two hourly periods before the baseline period are used to perform the parallel test. The coefficient estimates for the difference-in-differences terms in the hourly periods before the baseline period are used to test the parallel trend. Here, the coefficient estimates for $\var{Display} \times \var{Before:4}$ ($\var{coef.} =$ \num{0.015}, $p=$\num{0.142}) and $\var{Display} \times \var{Before:3}$ ($\var{coef.} =$ \num{0.015}, $p=$\num{0.148}) are not statistically significant. This suggests that the extra changes in reposts in the treatment group relative to the control group during the two before-display periods have no significant difference with that during the baseline period, which supports parallel trend. The coefficient estimates for the difference-in-differences terms in the hourly periods after the baseline period are used to estimate hourly multi-period ATTs.

\begin{table}
\centering
\caption{Regression results for parallel test and multi-period ATTs estimation. Post-specific random effects are included. Reported are coefficient estimates with standard errors in parentheses. \sym{*} \(p<0.01\), \sym{**} \(p<0.005\), \sym{***} \(p<0.001\).}
\resizebox{\textwidth}{!}{
\begin{adjustbox}{valign=t}
\begin{tabular}{l*{1}{S}}
\toprule
&\multicolumn{1}{c}{Multi-period}\\
\midrule
Display      &       1.279\sym{***}\\
            &     (0.024)         \\
Before:4     &       0.131\sym{***}\\
            &     (0.009)         \\
Before:3     &       0.036\sym{***}\\
            &     (0.008)         \\
After:1     &      -0.022         \\
            &     (0.009)         \\
After:2     &      -0.046\sym{***}\\
            &     (0.010)         \\
After:3     &      -0.060\sym{***}\\
            &     (0.011)         \\
After:4     &      -0.063\sym{***}\\
            &     (0.013)         \\
After:5     &      -0.066\sym{***}\\
            &     (0.015)         \\
After:6     &      -0.075\sym{***}\\
            &     (0.016)         \\
After:7     &      -0.065\sym{***}\\
            &     (0.018)         \\
After:8     &      -0.067\sym{***}\\
            &     (0.020)         \\
After:9     &      -0.059\sym{*}  \\
            &     (0.022)         \\
After:10    &      -0.045         \\
            &     (0.024)         \\
After:11    &      -0.038         \\
            &     (0.026)         \\
After:12    &      -0.023         \\
            &     (0.027)         \\
\bottomrule
\end{tabular}
\end{adjustbox}

\begin{adjustbox}{valign=t}
\begin{tabular}{l*{1}{S}}
\toprule
&\multicolumn{1}{c}{Continued}\\
\midrule
Display$\times$Before:4&       0.015         \\
            &     (0.010)         \\
Display$\times$Before:3&       0.015         \\
            &     (0.010)         \\
Display$\times$After:1&      -0.487\sym{***}\\
            &     (0.010)         \\
Display$\times$After:2&      -0.794\sym{***}\\
            &     (0.010)         \\
Display$\times$After:3&      -0.941\sym{***}\\
            &     (0.011)         \\
Display$\times$After:4&      -0.991\sym{***}\\
            &     (0.011)         \\
Display$\times$After:5&      -1.030\sym{***}\\
            &     (0.011)         \\
Display$\times$After:6&      -1.048\sym{***}\\
            &     (0.011)         \\
Display$\times$After:7&      -1.068\sym{***}\\
            &     (0.011)         \\
Display$\times$After:8&      -1.075\sym{***}\\
            &     (0.011)         \\
Display$\times$After:9&      -1.096\sym{***}\\
            &     (0.011)         \\
Display$\times$After:10&      -1.121\sym{***}\\
            &     (0.011)         \\
Display$\times$After:11&      -1.130\sym{***}\\
            &     (0.011)         \\
Display$\times$After:12&      -1.147\sym{***}\\
            &     (0.011)         \\
PostAge     &      -0.621\sym{***}\\
            &     (0.015)         \\
\bottomrule
\end{tabular}
\end{adjustbox}

\begin{adjustbox}{valign=t}
\begin{tabular}{l*{1}{S}}
\toprule
&\multicolumn{1}{c}{Continued}\\
\midrule
Intercept      &       1.322\sym{***}\\
            &     (0.021)         \\
\midrule
Post-level RE & \checkmark \\
\midrule
\#Observations       &      \num{614520}         \\
\#Posts       &      \num{40968}         \\
\bottomrule
\end{tabular}
\end{adjustbox}}
\label{tab:did_main_multi}
\end{table}

\newpage
\section{Placebo Analyses}
\label{supp:placebo}

We performed one-to-one matching again for the source posts in the treatment group and constructed a placebo group from source posts without displayed notes based on their propensity scores (see \Cref{supp:group_construction}). The placebo group demonstrates no statistically significant difference with the treatment group across the variables from user profiles and post features (\Cref{tab:psm_rollout}). We considered the placebo group as a virtual treatment group that is similar to the treatment group but did not receive actual treatment (\ie, note display). Subsequently, we investigated whether there was an additional reduction in reposts for placebo group during the after-display period compared to control group and relative to the before-display period. As shown in \Cref{fig:placebo}a, the changes in reposts within the placebo group overlapped with those within the control group. Based on the results from the regression that estimates multi-period ATTs using placebo and control groups (see \Cref{tab:did_placebo_multi_replace}), we found that all the hourly ATT estimations, whether before or after the note display, were not statistically significant. As visualized in \Cref{fig:placebo}b, there is no significant extra reductions after the display of community notes. Additionally, the two-period ATT estimation with the replacement of the actual treatment group was not statistically significant (ATT: \num{0.011}; 99\% CI: [\num{-0.002}, \num{0.024}]; $p=\num{0.028}$; see \Cref{tab:did_placebo_fixed}). These results suggest that the spread of the source posts in placebo and control groups followed a parallel pattern regardless of the note display.

\begin{figure}
	\captionsetup[subfloat]{font={bf, small}, skip=0pt, singlelinecheck=false, labelformat=simple, position=top}
	\centering
    \subfloat[]{\includegraphics[width = \textwidth]{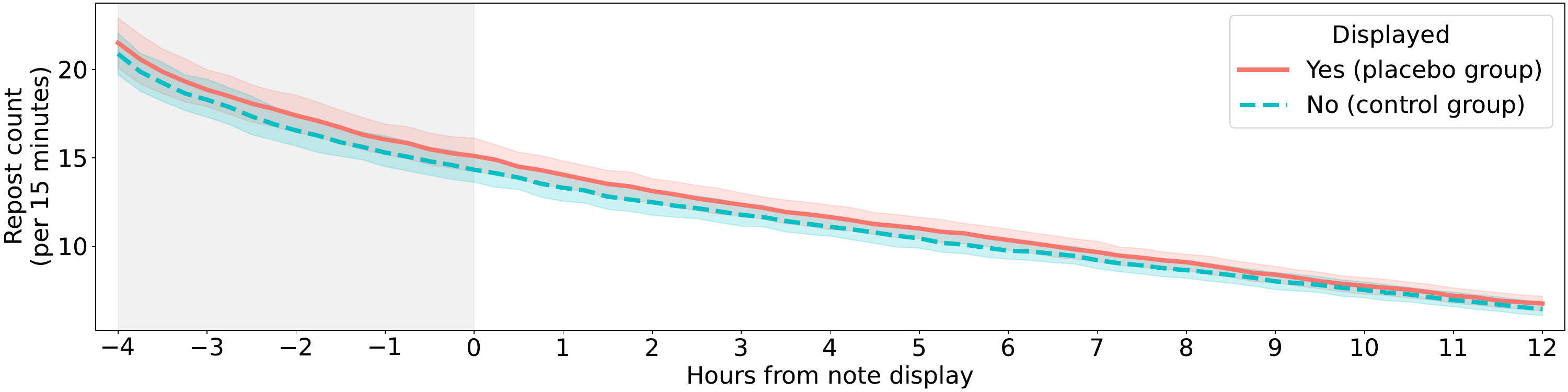}}
    \\
    \subfloat[]{\includegraphics[width = \textwidth]{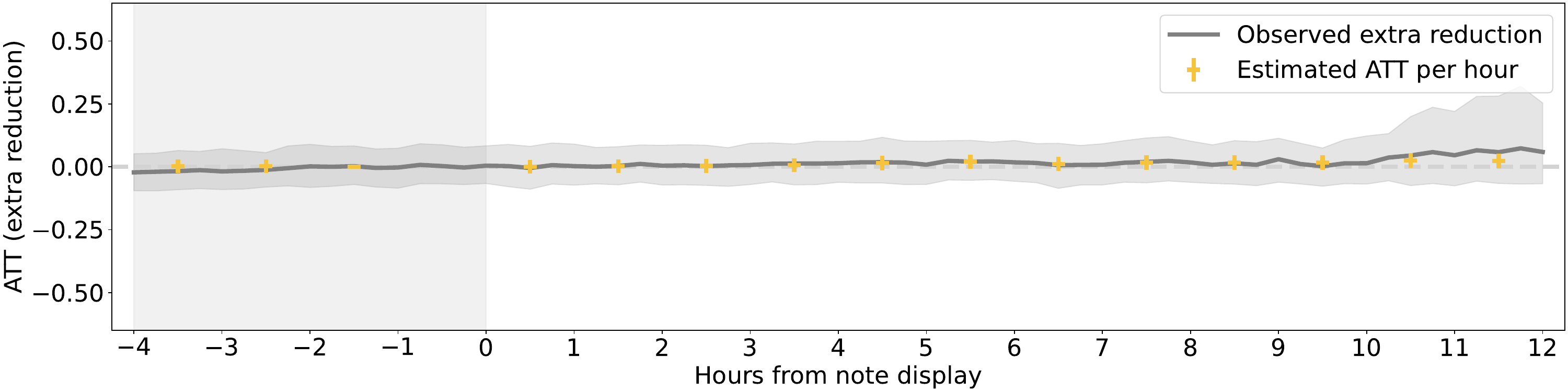}}
	\caption{\textbf{Placebo analyses for the effectiveness of community notes.}  \textbf{(a)}~The changes in repost count within the source posts in the placebo and control groups. The error bands represent 99\% CIs. \textbf{(b)}~The estimated hourly extra reductions (ATTs) of reposts in placebo group after the display of community notes, compared to control group and relative to before-display period (yellow). The error bars represent 99\% CIs. The grey belt (with 99\% CIs) visualizes the observed extra reduction of the ratio of reposts in the treatment group relative to reposts in the control group and compared to the ratio of reposts before the display of community notes.}
	\label{fig:placebo}
\end{figure}

\newpage
\begin{table}
\centering
\caption{Regression results for the multi-period ATT estimation with the replacement of treatment group. Post-specific random effects are included. Reported are coefficient estimates with standard errors in parentheses. \sym{*} \(p<0.01\), \sym{**} \(p<0.005\), \sym{***} \(p<0.001\).}
\resizebox{\textwidth}{!}{
\begin{adjustbox}{valign=t}
\begin{tabular}{l*{1}{S}}
\toprule
&\multicolumn{1}{c}{Multi-period}\\
\midrule
Display     &       0.034         \\
            &     (0.027)         \\
Before:4     &       0.101\sym{***}\\
            &     (0.009)         \\
Before:3     &       0.021\sym{*}  \\
            &     (0.008)         \\
After:1     &       0.011         \\
            &     (0.009)         \\
After:2     &       0.002         \\
            &     (0.010)         \\
After:3     &       0.003         \\
            &     (0.012)         \\
After:4     &       0.015         \\
            &     (0.014)         \\
After:5     &       0.027         \\
            &     (0.016)         \\
After:6     &       0.033         \\
            &     (0.018)         \\
After:7     &       0.058\sym{**} \\
            &     (0.020)         \\
After:8     &       0.071\sym{**} \\
            &     (0.022)         \\
After:9     &       0.094\sym{***}\\
            &     (0.024)         \\
After:10    &       0.122\sym{***}\\
            &     (0.026)         \\
After:11    &       0.144\sym{***}\\
            &     (0.028)         \\
After:12    &       0.174\sym{***}\\
            &     (0.030)         \\
\bottomrule
\end{tabular}
\end{adjustbox}

\begin{adjustbox}{valign=t}
\begin{tabular}{l*{1}{S}}
\toprule
&\multicolumn{1}{c}{Continued}\\
\midrule
Display$\times$Before:4&       0.003         \\
            &     (0.010)         \\
Display$\times$Before:3&       0.002         \\
            &     (0.010)         \\
Display$\times$After:1&       0.001         \\
            &     (0.011)         \\
Display$\times$After:2&       0.002         \\
            &     (0.011)         \\
Display$\times$After:3&       0.003         \\
            &     (0.011)         \\
Display$\times$After:4&       0.007         \\
            &     (0.011)         \\
Display$\times$After:5&       0.015         \\
            &     (0.011)         \\
Display$\times$After:6&       0.020         \\
            &     (0.011)         \\
Display$\times$After:7&       0.011         \\
            &     (0.011)         \\
Display$\times$After:8&       0.017         \\
            &     (0.011)         \\
Display$\times$After:9&       0.017         \\
            &     (0.011)         \\
Display$\times$After:10&       0.017         \\
            &     (0.011)         \\
Display$\times$After:11&       0.025         \\
            &     (0.011)         \\
Display$\times$After:12&       0.023         \\
            &     (0.011)         \\
PostAge     &      -0.731\sym{***}\\
            &     (0.016)         \\
\bottomrule
\end{tabular}
\end{adjustbox}

\begin{adjustbox}{valign=t}
\begin{tabular}{l*{1}{S}}
\toprule
&\multicolumn{1}{c}{Continued}\\
\midrule
Intercept      &       1.189\sym{***}\\
            &     (0.023)         \\
\midrule
Post-level RE & \checkmark\\
\midrule
\#Observations       &      \num{614520}         \\
\#Posts       &      \num{40968}         \\
\bottomrule
\end{tabular}
\end{adjustbox}}
\label{tab:did_placebo_multi_replace}
\end{table}

\begin{table}
\centering
\caption{Regression results for two-period ATT estimation with the replacement of treatment group. Post-specific random effects are included. Reported are coefficient estimates with standard errors in parentheses. \sym{*} \(p<0.01\), \sym{**} \(p<0.005\), \sym{***} \(p<0.001\).}
\resizebox{.43\textwidth}{!}{
\begin{tabular}{l*{2}{S}}
\toprule
&\multicolumn{1}{c}{(1)}           &\multicolumn{1}{c}{(2)}           \\
&\multicolumn{1}{c}{Main}&\multicolumn{1}{c}{Extended}\\
\midrule
Display     &       0.036         &       0.023         \\
            &     (0.026)         &     (0.025)         \\
After   &      -0.109\sym{***}&      -0.109\sym{***}\\
            &     (0.004)         &     (0.004)         \\
Display$\times$After&       0.011         &       0.011         \\
            &     (0.005)         &     (0.005)         \\
PostAge     &      -0.620\sym{***}&      -0.620\sym{***}\\
            &     (0.002)         &     (0.002)         \\
Verified    &                     &       1.484\sym{***}\\
            &                     &     (0.029)         \\
AccountAge  &                     &      -0.055\sym{***}\\
            &                     &     (0.013)         \\
Followers   &                     &       0.399\sym{***}\\
            &                     &     (0.018)         \\
Followees   &                     &       0.060\sym{***}\\
            &                     &     (0.012)         \\
Words       &                     &       0.088\sym{***}\\
            &                     &     (0.013)         \\
Media       &                     &       0.875\sym{***}\\
            &                     &     (0.028)         \\
Economy     &                     &      -0.376\sym{***}\\
            &                     &     (0.038)         \\
Health      &                     &       0.343\sym{***}\\
            &                     &     (0.042)         \\
Politics    &                     &       0.325\sym{***}\\
            &                     &     (0.030)         \\
Science     &                     &      -0.647\sym{***}\\
            &                     &     (0.037)         \\
Positive    &                     &      -0.128\sym{***}\\
            &                     &     (0.015)         \\
Negative    &                     &       0.289\sym{***}\\
            &                     &     (0.016)         \\
MFRO        &                     &      -0.046\sym{***}\\
            &                     &     (0.012)         \\
Intercept      &       1.337\sym{***}&      -0.336\sym{***}\\
            &     (0.019)         &     (0.036)         \\
\midrule
Post-level RE & \checkmark& \checkmark \\
\midrule
\#Observations       &      \num{614520}   &     \num{614520}\\
\#Posts       &      \num{40968}    &      \num{40968}\\
\bottomrule
\end{tabular}}
\label{tab:did_placebo_fixed}
\end{table}

\newpage
\section{Sensitivity Analyses}
\label{supp:sensitivity}

\subsection{Sensitivity Across Response Time}

We analyzed how the efficacy of community notes varies depending on the post age at note display, \ie, the response time. We conducted the regressions across the source posts grouped by the response time with a window of 4 hours. For instance, the source posts that received displayed notes at \num{4} and \num{8} hours from post creation are grouped together for the subset analysis. The regression results are reported in \Cref{tab:did_sensitivity_cohorts}.

\begin{table}
\centering
\caption{Regression results for two-period ATT estimations over response time, \ie, post age at note display from 4 to 24 hours. Post-specific random effects are included. Reported are coefficient estimates with standard errors in parentheses. \sym{*} \(p<0.01\), \sym{**} \(p<0.005\), \sym{***} \(p<0.001\).}
\begin{subtable}{\textwidth}
\resizebox{\textwidth}{!}{
\begin{tabular}{l*{5}{S}}
\toprule
Response time &{4~--~8}&{8~--~12}&{12~--~16}&{16~--~20}&{20~--~24}\\
\midrule
Display&       1.281\sym{***}&       1.192\sym{***}&       1.251\sym{***}&       1.427\sym{***}&       1.318\sym{***}\\
            &     (0.040)         &     (0.050)         &     (0.058)         &     (0.062)         &     (0.064)         \\
After&      -0.205\sym{***}&       0.014         &       0.346\sym{***}&       0.781\sym{***}&       0.226\sym{***}\\
            &     (0.008)         &     (0.009)         &     (0.011)         &     (0.012)         &     (0.012)         \\
Display $\times$ After&      -1.141\sym{***}&      -0.978\sym{***}&      -0.953\sym{***}&      -0.830\sym{***}&      -0.617\sym{***}\\
            &     (0.009)         &     (0.010)         &     (0.011)         &     (0.012)         &     (0.013)         \\
PostAge&      -0.641\sym{***}&      -0.499\sym{***}&      -0.757\sym{***}&      -1.442\sym{***}&      -1.433\sym{***}\\
            &     (0.004)         &     (0.005)         &     (0.006)         &     (0.007)         &     (0.007)         \\
Intercept&       1.476\sym{***}&       1.356\sym{***}&       1.138\sym{***}&       1.128\sym{***}&       1.827\sym{***}\\
            &     (0.029)         &     (0.036)         &     (0.041)         &     (0.043)         &     (0.045)         \\
\midrule
Post-level RE & \checkmark& \checkmark & \checkmark& \checkmark& \checkmark\\
\midrule
\#Observations       &       \num{197370}         &       \num{136620}         &       \num{102150}         &       \num{92430}         &       \num{85950}         \\
\bottomrule
\end{tabular}}
\end{subtable}
\label{tab:did_sensitivity_cohorts}
\end{table}

\newpage
\subsection{Sensitivity Across Months From Roll-Out}

The regression results for the two-period ATT estimation model across months from the roll-out (MFRO) of ``Community Notes'' program are reported in \Cref{tab:did_sensitivity_mfro}.

\begin{table}
\centering
\caption{Regression results for two-period ATT estimations across months from the roll-out of ``Community Notes'' program. Post-specific random effects are included. Reported are coefficient estimates with standard errors in parentheses. \sym{*} \(p<0.01\), \sym{**} \(p<0.005\), \sym{***} \(p<0.001\).}
\begin{subtable}{\textwidth}
\resizebox{\textwidth}{!}{
\begin{tabular}{l*{7}{S}}
\toprule
MFRO&{1}&{2}&{3}&{4}&{5}&{6}&{7}\\
\midrule
Display     &       1.576\sym{***}&       1.465\sym{***}&       1.115\sym{***}&       1.827\sym{***}&       1.429\sym{***}&       0.993\sym{***}&       1.481\sym{***}\\
            &     (0.382)         &     (0.290)         &     (0.277)         &     (0.277)         &     (0.216)         &     (0.173)         &     (0.144)         \\
After   &      -0.073         &      -0.139\sym{**} &       0.026         &       0.189\sym{***}&       0.196\sym{***}&       0.151\sym{***}&       0.061         \\
            &     (0.054)         &     (0.042)         &     (0.043)         &     (0.044)         &     (0.038)         &     (0.029)         &     (0.026)         \\
Display $\times$ After&      -0.280\sym{***}&      -0.285\sym{***}&      -0.551\sym{***}&      -0.647\sym{***}&      -0.804\sym{***}&      -0.686\sym{***}&      -0.609\sym{***}\\
            &     (0.061)         &     (0.047)         &     (0.045)         &     (0.049)         &     (0.041)         &     (0.033)         &     (0.029)         \\
PostAge     &      -0.784\sym{***}&      -0.749\sym{***}&      -0.804\sym{***}&      -0.779\sym{***}&      -0.808\sym{***}&      -0.876\sym{***}&      -0.867\sym{***}\\
            &     (0.030)         &     (0.023)         &     (0.022)         &     (0.024)         &     (0.021)         &     (0.016)         &     (0.015)         \\
Intercept      &       1.478\sym{***}&       2.111\sym{***}&       1.725\sym{***}&       1.345\sym{***}&       1.206\sym{***}&       1.411\sym{***}&       1.159\sym{***}\\
            &     (0.243)         &     (0.194)         &     (0.206)         &     (0.181)         &     (0.145)         &     (0.114)         &     (0.095)         \\
\midrule
Post-level RE & \checkmark& \checkmark & \checkmark& \checkmark& \checkmark& \checkmark& \checkmark\\
\midrule
\#Observations       &      \num{3315}         &      \num{5115}         &      \num{5280}         &      \num{5520}         &      \num{8265}         &      \num{11955}         &      \num{15735}\\
\bottomrule
\end{tabular}}
\end{subtable}

\vspace{1em}

\begin{subtable}{\textwidth}
\resizebox{\textwidth}{!}{
\begin{tabular}{l*{7}{S}}
\toprule
MFRO&{8}&{9}&{10}&{11}&{12}&{13}&{14}\\
\midrule
Display     &       1.346\sym{***}&       1.391\sym{***}&       1.746\sym{***}&       1.581\sym{***}&       1.625\sym{***}&       1.509\sym{***}&       1.149\sym{***}\\
            &     (0.122)         &     (0.122)         &     (0.122)         &     (0.097)         &     (0.109)         &     (0.085)         &     (0.088)         \\
After   &       0.165\sym{***}&       0.162\sym{***}&       0.249\sym{***}&       0.411\sym{***}&       0.297\sym{***}&       0.062\sym{***}&       0.085\sym{***}\\
            &     (0.023)         &     (0.024)         &     (0.023)         &     (0.020)         &     (0.022)         &     (0.016)         &     (0.016)         \\
Display $\times$ After&      -0.645\sym{***}&      -0.664\sym{***}&      -0.666\sym{***}&      -0.816\sym{***}&      -0.771\sym{***}&      -0.984\sym{***}&      -0.976\sym{***}\\
            &     (0.025)         &     (0.025)         &     (0.025)         &     (0.021)         &     (0.023)         &     (0.017)         &     (0.018)         \\
PostAge     &      -0.928\sym{***}&      -0.883\sym{***}&      -0.885\sym{***}&      -0.950\sym{***}&      -0.905\sym{***}&      -0.857\sym{***}&      -0.851\sym{***}\\
            &     (0.013)         &     (0.013)         &     (0.012)         &     (0.010)         &     (0.012)         &     (0.008)         &     (0.009)         \\
Intercept      &       1.002\sym{***}&       1.138\sym{***}&       1.136\sym{***}&       0.968\sym{***}&       0.948\sym{***}&       1.136\sym{***}&       1.293\sym{***}\\
            &     (0.085)         &     (0.086)         &     (0.083)         &     (0.070)         &     (0.076)         &     (0.061)         &     (0.062)         \\
\midrule
Post-level RE & \checkmark& \checkmark & \checkmark& \checkmark& \checkmark& \checkmark& \checkmark\\
\midrule
\#Observations       &      \num{20955}         &      \num{20700}         &      \num{22920}         &      \num{34830}         &      \num{25335}         &      \num{48615}         &      \num{45480}\\
\bottomrule
\end{tabular}}
\end{subtable}

\vspace{1em}

\begin{subtable}{\textwidth}
\resizebox{\textwidth}{!}{
\begin{tabular}{l*{7}{S}}
\toprule
MFRO&{15}&{16}&{17}&{18}&{19}&{20}&{21}\\
\midrule
Display     &       0.916\sym{***}&       1.345\sym{***}&       1.263\sym{***}&       1.225\sym{***}&       1.157\sym{***}&       1.168\sym{***}&       1.328\sym{***}\\
            &     (0.080)         &     (0.081)         &     (0.090)         &     (0.075)         &     (0.074)         &     (0.073)         &     (0.127)         \\
After   &       0.078\sym{***}&       0.010         &       0.065\sym{***}&       0.092\sym{***}&       0.028         &      -0.003         &      -0.008         \\
            &     (0.015)         &     (0.016)         &     (0.017)         &     (0.016)         &     (0.016)         &     (0.017)         &     (0.028)         \\
Display $\times$ After&      -1.043\sym{***}&      -1.017\sym{***}&      -1.123\sym{***}&      -1.114\sym{***}&      -1.101\sym{***}&      -1.197\sym{***}&      -1.220\sym{***}\\
            &     (0.016)         &     (0.017)         &     (0.019)         &     (0.017)         &     (0.017)         &     (0.018)         &     (0.030)         \\
PostAge     &      -0.720\sym{***}&      -0.749\sym{***}&      -0.751\sym{***}&      -0.770\sym{***}&      -0.765\sym{***}&      -0.725\sym{***}&      -0.711\sym{***}\\
            &     (0.008)         &     (0.009)         &     (0.009)         &     (0.009)         &     (0.009)         &     (0.009)         &     (0.015)         \\
Intercept      &       1.395\sym{***}&       1.193\sym{***}&       1.329\sym{***}&       1.088\sym{***}&       1.213\sym{***}&       1.220\sym{***}&       1.175\sym{***}\\
            &     (0.058)         &     (0.056)         &     (0.063)         &     (0.057)         &     (0.056)         &     (0.054)         &     (0.090)         \\
\midrule
Post-level RE & \checkmark& \checkmark & \checkmark& \checkmark& \checkmark& \checkmark& \checkmark\\
\midrule
\#Observations       &      \num{53205}         &      \num{55170}         &      \num{44400}         &      \num{55695}         &      \num{56445}         &      \num{54690}         &      \num{20895}\\
\bottomrule
\end{tabular}}
\end{subtable}
\label{tab:did_sensitivity_mfro}
\end{table}

\newpage
\subsection{Sensitivity Across Rating Thresholds}

The regression results for the two-period ATT estimation model under different rating thresholds are reported in \Cref{tab:did_sensitivity_ratings}. Here, we restricted our analysis to the source posts with community notes that had reached a specific threshold for the number of ratings. 
Here, we only considered posts with community notes that indicated misleading had never been rated as not helpful. 
The rating thresholds in our sensitivity analysis range from 10 to 80 with a step size of 10.

\begin{table}
\centering
\caption{Regression results for two-period ATT estimations across rating thresholds. Post-specific random effects are included. Reported are coefficient estimates with standard errors in parentheses. \sym{*} \(p<0.01\), \sym{**} \(p<0.005\), \sym{***} \(p<0.001\).}
\begin{subtable}{\textwidth}
\resizebox{\textwidth}{!}{
\begin{tabular}{l*{8}{S}}
\toprule
Ratings&{10}&{20}&{30}&{40}&{50}&{60}&{70}&{80}\\
\midrule
Display     &       0.237\sym{***}&      -0.151\sym{***}&      -0.342\sym{***}&      -0.426\sym{***}&      -0.483\sym{***}&      -0.499\sym{***}&      -0.505\sym{***}&      -0.502\sym{***}\\
            &     (0.026)         &     (0.028)         &     (0.029)         &     (0.031)         &     (0.033)         &     (0.035)         &     (0.036)         &     (0.038)         \\
After   &       0.152\sym{***}&       0.185\sym{***}&       0.199\sym{***}&       0.212\sym{***}&       0.217\sym{***}&       0.220\sym{***}&       0.221\sym{***}&       0.217\sym{***}\\
            &     (0.006)         &     (0.006)         &     (0.007)         &     (0.007)         &     (0.008)         &     (0.008)         &     (0.008)         &     (0.009)         \\
Display $\times$ After&      -0.988\sym{***}&      -1.004\sym{***}&      -1.008\sym{***}&      -1.013\sym{***}&      -1.011\sym{***}&      -1.011\sym{***}&      -1.004\sym{***}&      -0.996\sym{***}\\
            &     (0.006)         &     (0.006)         &     (0.007)         &     (0.007)         &     (0.007)         &     (0.008)         &     (0.008)         &     (0.009)         \\
PostAge     &      -0.837\sym{***}&      -0.853\sym{***}&      -0.862\sym{***}&      -0.868\sym{***}&      -0.870\sym{***}&      -0.868\sym{***}&      -0.864\sym{***}&      -0.855\sym{***}\\
            &     (0.003)         &     (0.003)         &     (0.003)         &     (0.003)         &     (0.003)         &     (0.003)         &     (0.003)         &     (0.003)         \\
Intercept      &       2.233\sym{***}&       2.639\sym{***}&       2.885\sym{***}&       3.051\sym{***}&       3.191\sym{***}&       3.285\sym{***}&       3.376\sym{***}&       3.454\sym{***}\\
            &     (0.021)         &     (0.024)         &     (0.026)         &     (0.028)         &     (0.029)         &     (0.031)         &     (0.033)         &     (0.034)         \\
\midrule
Post-level RE & \checkmark& \checkmark & \checkmark& \checkmark& \checkmark& \checkmark& \checkmark& \checkmark\\
\midrule
\#Observations       &      \num{465600}         &      \num{419940}         &      \num{387360}         &      \num{357810}         &      \num{328065}         &      \num{302490}         &      \num{277740}         &      \num{253650}\\
\bottomrule
\end{tabular}}
\end{subtable}
\label{tab:did_sensitivity_ratings}
\end{table}

\newpage
\subsection{Sensitivity Across User and Post Characteristics}

\Cref{tab:did_sensitivity_user_charcs} reports the regression results for the sensitivity analysis on how the efficacy of community notes varied depending on user characteristics ($\var{Verified}$, $\var{AccountAge}$, $\var{Followers}$, and $\var{Followees}$).
\Cref{tab:did_sensitivity_post_charcs1} and \Cref{tab:did_sensitivity_post_charcs2} report the regression results for the sensitivity analysis on how the efficacy of community notes varied depending on post characteristics ($\var{Words}$, $\var{Media}$, $\var{Positive}$, $\var{Negative}$, $\var{Economy}$, $\var{Health}$, $\var{Politics}$, and $\var{Science}$).

\begin{table}
\centering
\caption{Regression results for two-period ATT estimations across user characteristics, \ie, $\var{Verified}$, $\var{AccountAge}$, $\var{Followers}$ and $\var{Followees}$. The estimation models incorporate interaction terms and subsets separately. Post-specific random effects are included. Reported are coefficient estimates with standard errors in parentheses. \sym{*} \(p<0.01\), \sym{**} \(p<0.005\), \sym{***} \(p<0.001\).}
\begin{subtable}{\textwidth}
\resizebox{\textwidth}{!}{
\begin{tabular}{l*{4}{S}}
\toprule
&{Interaction}&{Verified}&{Not verified}\\
\midrule
Display     &       1.986\sym{***}&       1.049\sym{***}&       2.015\sym{***}\\
            &     (0.049)         &     (0.025)         &     (0.054)         \\
After   &       0.217\sym{***}&       0.076\sym{***}&       0.202\sym{***}\\
            &     (0.009)         &     (0.005)         &     (0.013)         \\
Display$\times$After&      -1.345\sym{***}&      -0.889\sym{***}&      -1.338\sym{***}\\
            &     (0.012)         &     (0.005)         &     (0.014)         \\
PostAge     &      -0.807\sym{***}&      -0.806\sym{***}&      -0.812\sym{***}\\
            &     (0.003)         &     (0.003)         &     (0.007)         \\
Verified    &       1.692\sym{***}&                     &                     \\
            &     (0.038)         &                     &                     \\
Display$\times$Verified&      -0.933\sym{***}&                     &                     \\
            &     (0.055)         &                     &                     \\
After$\times$Verified&      -0.143\sym{***}&                     &                     \\
            &     (0.010)         &                     &                     \\
ATT$\times$Verified&       0.456\sym{***}&                     &                     \\
            &     (0.013)         &                     &                     \\
Intercept      &      -0.097\sym{**} &       1.599\sym{***}&      -0.130\sym{***}\\
            &     (0.033)         &     (0.018)         &     (0.037)         \\
\midrule
Post-level RE & \checkmark& \checkmark& \checkmark \\
\midrule
\#Observations       &      \num{614520}         &      \num{478170}         &      \num{136350}         \\
\bottomrule
\end{tabular}

\begin{tabular}{l*{4}{S}}
\toprule
&{Interaction}&{High account age}&{Low account age}\\
\midrule
Display     &       1.260\sym{***}&       1.113\sym{***}&       1.461\sym{***}\\
            &     (0.024)         &     (0.032)         &     (0.034)         \\
After   &       0.095\sym{***}&       0.060\sym{***}&       0.139\sym{***}\\
            &     (0.005)         &     (0.006)         &     (0.007)         \\
Display$\times$After&      -0.954\sym{***}&      -0.877\sym{***}&      -1.061\sym{***}\\
            &     (0.005)         &     (0.007)         &     (0.007)         \\
PostAge     &      -0.807\sym{***}&      -0.803\sym{***}&      -0.811\sym{***}\\
            &     (0.003)         &     (0.004)         &     (0.004)         \\
AccountAge  &       0.091\sym{***}&                     &                     \\
            &     (0.017)         &                     &                     \\
Display$\times$AccountAge&      -0.227\sym{***}&                     &                     \\
            &     (0.024)         &                     &                     \\
After$\times$AccountAge&      -0.047\sym{***}&                     &                     \\
            &     (0.004)         &                     &                     \\
ATT$\times$AccountAge&       0.124\sym{***}&                     &                     \\
            &     (0.005)         &                     &                     \\
Intercept      &       1.211\sym{***}&       1.271\sym{***}&       1.133\sym{***}\\
            &     (0.017)         &     (0.023)         &     (0.024)         \\
\midrule
Post-level RE & \checkmark& \checkmark& \checkmark \\
\midrule
\#Observations       &      \num{614520}         &      \num{307200}         &      \num{307320}         \\
\bottomrule
\end{tabular}}
\end{subtable}

\vspace{1em}

\begin{subtable}{\textwidth}
\resizebox{\textwidth}{!}{
\begin{tabular}{l*{4}{S}}
\toprule
&{Interaction}&{High followers}&{Low followers}\\
\midrule
Display      &       1.278\sym{***}&       0.582\sym{***}&       1.780\sym{***}\\
            &     (0.023)         &     (0.029)         &     (0.033)         \\
After   &       0.098\sym{***}&       0.069\sym{***}&       0.130\sym{***}\\
            &     (0.005)         &     (0.006)         &     (0.008)         \\
Display$\times$After&      -0.963\sym{***}&      -0.816\sym{***}&      -1.179\sym{***}\\
            &     (0.005)         &     (0.006)         &     (0.009)         \\
PostAge     &      -0.807\sym{***}&      -0.815\sym{***}&      -0.790\sym{***}\\
            &     (0.003)         &     (0.003)         &     (0.004)         \\
Followers   &       0.409\sym{***}&                     &                     \\
            &     (0.024)         &                     &                     \\
Display$\times$Followers&      -0.199\sym{***}&                     &                     \\
            &     (0.037)         &                     &                     \\
After$\times$Followers&      -0.023\sym{***}&                     &                     \\
            &     (0.005)         &                     &                     \\
ATT$\times$Followers&       0.109\sym{***}&                     &                     \\
            &     (0.007)         &                     &                     \\
Intercept      &       1.224\sym{***}&       2.381\sym{***}&       0.180\sym{***}\\
            &     (0.017)         &     (0.022)         &     (0.023)         \\
\midrule
Post-level RE & \checkmark& \checkmark& \checkmark \\
\midrule
\#Observations       &      \num{614520}         &      \num{305610}         &      \num{308910}         \\
\bottomrule
\end{tabular}

\begin{tabular}{l*{4}{S}}
\toprule
&{Interaction}&{High followees}&{Low followees}\\
\midrule
Display     &       1.280\sym{***}&       1.245\sym{***}&       1.315\sym{***}\\
            &     (0.023)         &     (0.032)         &     (0.035)         \\
After   &       0.099\sym{***}&       0.113\sym{***}&       0.083\sym{***}\\
            &     (0.005)         &     (0.007)         &     (0.007)         \\
Display$\times$After&      -0.967\sym{***}&      -1.007\sym{***}&      -0.924\sym{***}\\
            &     (0.005)         &     (0.007)         &     (0.007)         \\
PostAge     &      -0.807\sym{***}&      -0.831\sym{***}&      -0.782\sym{***}\\
            &     (0.003)         &     (0.004)         &     (0.004)         \\
Followees   &       0.137\sym{***}&                     &                     \\
            &     (0.016)         &                     &                     \\
Display$\times$Followees&       0.068\sym{*}  &                     &                     \\
            &     (0.024)         &                     &                     \\
After$\times$Followees&      -0.004         &                     &                     \\
            &     (0.003)         &                     &                     \\
ATT$\times$Followees&      -0.031\sym{***}&                     &                     \\
            &     (0.005)         &                     &                     \\
Intercept      &       1.203\sym{***}&       1.342\sym{***}&       1.066\sym{***}\\
            &     (0.017)         &     (0.023)         &     (0.024)         \\
\midrule
Post-level RE & \checkmark& \checkmark& \checkmark \\
\midrule
\#Observations       &      \num{614520}         &      \num{307125}         &      \num{307395}         \\
\bottomrule
\end{tabular}}
\end{subtable}
\label{tab:did_sensitivity_user_charcs}
\end{table}

\begin{table}
\centering
\caption{Regression results for two-period ATT estimations across post characteristics including $\var{Words}$, $\var{Media}$, $\var{Positive}$ and $\var{Negative}$. The estimation models incorporate interaction terms and subsets separately. Post-specific random effects are included. Reported are coefficient estimates with standard errors in parentheses. \sym{*} \(p<0.01\), \sym{**} \(p<0.005\), \sym{***} \(p<0.001\).}
\begin{subtable}{\textwidth}
\resizebox{\textwidth}{!}{
\begin{tabular}{l*{4}{S}}
\toprule
&{Interaction}&{High words}&{Low words}\\
\midrule
Display     &       1.254\sym{***}&       1.100\sym{***}&       1.479\sym{***}\\
            &     (0.024)         &     (0.032)         &     (0.034)         \\
After   &       0.096\sym{***}&       0.052\sym{***}&       0.150\sym{***}\\
            &     (0.005)         &     (0.006)         &     (0.007)         \\
Display$\times$After&      -0.955\sym{***}&      -0.893\sym{***}&      -1.046\sym{***}\\
            &     (0.005)         &     (0.007)         &     (0.007)         \\
PostAge     &      -0.807\sym{***}&      -0.787\sym{***}&      -0.828\sym{***}\\
            &     (0.003)         &     (0.003)         &     (0.004)         \\
Words       &       0.207\sym{***}&                     &                     \\
            &     (0.017)         &                     &                     \\
Display$\times$Words&      -0.205\sym{***}&                     &                     \\
            &     (0.024)         &                     &                     \\
After$\times$Words&      -0.022\sym{***}&                     &                     \\
            &     (0.004)         &                     &                     \\
ATT$\times$Words   &       0.084\sym{***}&                     &                     \\
            &     (0.005)         &                     &                     \\
Intercept      &       1.235\sym{***}&       1.386\sym{***}&       1.011\sym{***}\\
            &     (0.017)         &     (0.023)         &     (0.024)         \\
\midrule
Post-level RE & \checkmark& \checkmark& \checkmark \\
\midrule
\#Observations       &      \num{614520}         &      \num{305910}         &      \num{308610}         \\
\bottomrule
\end{tabular}

\begin{tabular}{l*{4}{S}}
\toprule
&{Interaction}&{Media}&{No media}\\
\midrule
Display     &       1.175\sym{***}&       1.295\sym{***}&       1.193\sym{***}\\
            &     (0.044)         &     (0.027)         &     (0.047)         \\
After   &       0.024\sym{**} &       0.134\sym{***}&      -0.002         \\
            &     (0.008)         &     (0.005)         &     (0.009)         \\
Display$\times$After&      -0.776\sym{***}&      -1.032\sym{***}&      -0.775\sym{***}\\
            &     (0.010)         &     (0.006)         &     (0.010)         \\
PostAge     &      -0.807\sym{***}&      -0.814\sym{***}&      -0.788\sym{***}\\
            &     (0.003)         &     (0.003)         &     (0.005)         \\
Media       &       0.593\sym{***}&                     &                     \\
            &     (0.036)         &                     &                     \\
Display$\times$Media&       0.125         &                     &                     \\
            &     (0.052)         &                     &                     \\
After$\times$Media&       0.102\sym{***}&                     &                     \\
            &     (0.008)         &                     &                     \\
ATT$\times$Media   &      -0.256\sym{***}&                     &                     \\
            &     (0.011)         &                     &                     \\
Intercept      &       0.782\sym{***}&       1.375\sym{***}&       0.775\sym{***}\\
            &     (0.031)         &     (0.019)         &     (0.033)         \\
\midrule
Post-level RE & \checkmark& \checkmark& \checkmark \\
\midrule
\#Observations       &      \num{614520}         &      \num{444360}         &      \num{170160}         \\
\bottomrule
\end{tabular}}
\end{subtable}

\vspace{1em}

\begin{subtable}{\textwidth}
\resizebox{\textwidth}{!}{
\begin{tabular}{l*{4}{S}}
\toprule
&{Interaction}&{High positive}&{Low positive}\\
\midrule
Display      &       1.271\sym{***}&       1.500\sym{***}&       1.066\sym{***}\\
            &     (0.023)         &     (0.033)         &     (0.033)         \\
After   &       0.099\sym{***}&       0.107\sym{***}&       0.091\sym{***}\\
            &     (0.005)         &     (0.007)         &     (0.006)         \\
Display$\times$After&      -0.968\sym{***}&      -0.974\sym{***}&      -0.964\sym{***}\\
            &     (0.005)         &     (0.007)         &     (0.007)         \\
PostAge     &      -0.807\sym{***}&      -0.803\sym{***}&      -0.811\sym{***}\\
            &     (0.003)         &     (0.004)         &     (0.003)         \\
Positive    &      -0.362\sym{***}&                     &                     \\
            &     (0.016)         &                     &                     \\
Display$\times$Positive&       0.193\sym{***}&                     &                     \\
            &     (0.023)         &                     &                     \\
After$\times$Positive&       0.003         &                     &                     \\
            &     (0.004)         &                     &                     \\
ATT$\times$Positive&       0.009         &                     &                     \\
            &     (0.005)         &                     &                     \\
Intercept      &       1.217\sym{***}&       0.871\sym{***}&       1.539\sym{***}\\
            &     (0.017)         &     (0.024)         &     (0.023)         \\
\midrule
Post-level RE & \checkmark& \checkmark& \checkmark \\
\midrule
\#Observations       &      \num{614520}         &      \num{307260}         &      \num{307260}         \\
\bottomrule
\end{tabular}

\begin{tabular}{l*{4}{S}}
\toprule
&{Interaction}&{High negative}&{Low negative}\\
\midrule
Display     &       1.261\sym{***}&       1.043\sym{***}&       1.522\sym{***}\\
            &     (0.023)         &     (0.033)         &     (0.033)         \\
After   &       0.098\sym{***}&       0.088\sym{***}&       0.110\sym{***}\\
            &     (0.005)         &     (0.006)         &     (0.007)         \\
Display$\times$After&      -0.969\sym{***}&      -0.964\sym{***}&      -0.974\sym{***}\\
            &     (0.005)         &     (0.007)         &     (0.007)         \\
PostAge     &      -0.807\sym{***}&      -0.804\sym{***}&      -0.810\sym{***}\\
            &     (0.003)         &     (0.003)         &     (0.004)         \\
Negative    &       0.401\sym{***}&                     &                     \\
            &     (0.016)         &                     &                     \\
Display$\times$Negative&      -0.257\sym{***}&                     &                     \\
            &     (0.023)         &                     &                     \\
After$\times$Negative&      -0.007         &                     &                     \\
            &     (0.004)         &                     &                     \\
ATT$\times$Negative&       0.003         &                     &                     \\
            &     (0.005)         &                     &                     \\
Intercept      &       1.240\sym{***}&       1.577\sym{***}&       0.833\sym{***}\\
            &     (0.017)         &     (0.023)         &     (0.024)         \\
\midrule
Post-level RE & \checkmark& \checkmark& \checkmark \\
\midrule
\#Observations       &      \num{614520}         &      \num{307260}         &      \num{307260}         \\
\bottomrule
\end{tabular}}
\end{subtable}
\label{tab:did_sensitivity_post_charcs1}
\end{table}

\begin{table}
\centering
\caption{Regression results for two-period ATT estimations across post characteristics including $\var{Economy}$, $\var{Health}$, $\var{Politics}$ and $\var{Science}$. The estimation models incorporate interaction terms and subsets separately. Post-specific random effects are included. Reported are coefficient estimates with standard errors in parentheses. \sym{*} \(p<0.01\), \sym{**} \(p<0.005\), \sym{***} \(p<0.001\).}
\begin{subtable}{\textwidth}
\resizebox{\textwidth}{!}{
\begin{tabular}{l*{4}{S}}
\toprule
&{Interaction}&{With economy}&{Without economy}\\
\midrule
Display     &       1.277\sym{***}&       1.331\sym{***}&       1.277\sym{***}\\
            &     (0.025)         &     (0.068)         &     (0.025)         \\
After   &       0.103\sym{***}&       0.073\sym{***}&       0.102\sym{***}\\
            &     (0.005)         &     (0.014)         &     (0.005)         \\
Display$\times$After&      -0.981\sym{***}&      -0.868\sym{***}&      -0.981\sym{***}\\
            &     (0.005)         &     (0.015)         &     (0.005)         \\
PostAge     &      -0.807\sym{***}&      -0.813\sym{***}&      -0.806\sym{***}\\
            &     (0.003)         &     (0.008)         &     (0.003)         \\
Economy     &      -0.467\sym{***}&                     &                     \\
            &     (0.050)         &                     &                     \\
Display$\times$Economy&       0.050         &                     &                     \\
            &     (0.072)         &                     &                     \\
After$\times$Economy&      -0.034\sym{**} &                     &                     \\
            &     (0.012)         &                     &                     \\
ATT$\times$Economy &       0.114\sym{***}&                     &                     \\
            &     (0.016)         &                     &                     \\
Intercept      &       1.261\sym{***}&       0.785\sym{***}&       1.262\sym{***}\\
            &     (0.018)         &     (0.048)         &     (0.018)         \\
\midrule
Post-level RE & \checkmark& \checkmark& \checkmark \\
\midrule
\#Observations       &      \num{614520}         &      \num{74340}         &      \num{540180}         \\
\bottomrule
\end{tabular}

\begin{tabular}{l*{4}{S}}
\toprule
&{Interaction}&{With health}&{Without health}\\
\midrule
Display     &       1.309\sym{***}&       1.063\sym{***}&       1.311\sym{***}\\
            &     (0.025)         &     (0.072)         &     (0.025)         \\
After   &       0.091\sym{***}&       0.025         &       0.108\sym{***}\\
            &     (0.005)         &     (0.013)         &     (0.005)         \\
Display$\times$After&      -0.983\sym{***}&      -0.821\sym{***}&      -0.984\sym{***}\\
            &     (0.005)         &     (0.014)         &     (0.005)         \\
PostAge     &      -0.807\sym{***}&      -0.681\sym{***}&      -0.823\sym{***}\\
            &     (0.003)         &     (0.007)         &     (0.003)         \\
Health      &       0.443\sym{***}&                     &                     \\
            &     (0.056)         &                     &                     \\
Display$\times$Health&      -0.235\sym{**} &                     &                     \\
            &     (0.080)         &                     &                     \\
After$\times$Health&       0.072\sym{***}&                     &                     \\
            &     (0.012)         &                     &                     \\
ATT$\times$Health  &       0.157\sym{***}&                     &                     \\
            &     (0.016)         &                     &                     \\
Intercept      &       1.160\sym{***}&       1.738\sym{***}&       1.144\sym{***}\\
            &     (0.018)         &     (0.050)         &     (0.018)         \\
\midrule
Post-level RE & \checkmark& \checkmark& \checkmark \\
\midrule
\#Observations       &      \num{614520}         &      \num{58245}         &      \num{556275}         \\
\bottomrule
\end{tabular}}
\end{subtable}

\vspace{1em}

\begin{subtable}{\textwidth}
\resizebox{\textwidth}{!}{
\begin{tabular}{l*{4}{S}}
\toprule
&{Interaction}&{With politics}&{With politics}\\
\midrule
Display      &       1.407\sym{***}&       0.945\sym{***}&       1.413\sym{***}\\
            &     (0.027)         &     (0.043)         &     (0.028)         \\
After   &       0.132\sym{***}&       0.021         &       0.130\sym{***}\\
            &     (0.005)         &     (0.009)         &     (0.005)         \\
Display$\times$After&      -1.011\sym{***}&      -0.862\sym{***}&      -1.011\sym{***}\\
            &     (0.006)         &     (0.009)         &     (0.006)         \\
PostAge     &      -0.807\sym{***}&      -0.809\sym{***}&      -0.806\sym{***}\\
            &     (0.003)         &     (0.005)         &     (0.003)         \\
Politics    &       0.661\sym{***}&                     &                     \\
            &     (0.038)         &                     &                     \\
Display$\times$Politics&      -0.452\sym{***}&                     &                     \\
            &     (0.054)         &                     &                     \\
After$\times$Politics&      -0.115\sym{***}&                     &                     \\
            &     (0.008)         &                     &                     \\
ATT$\times$Politics&       0.150\sym{***}&                     &                     \\
            &     (0.011)         &                     &                     \\
Intercept      &       1.029\sym{***}&       1.702\sym{***}&       1.022\sym{***}\\
            &     (0.019)         &     (0.031)         &     (0.020)         \\
\midrule
Post-level RE & \checkmark& \checkmark& \checkmark \\
\midrule
\#Observations       &      \num{614520}         &      \num{157635}         &      \num{456885}         \\
\bottomrule
\end{tabular}

\begin{tabular}{l*{4}{S}}
\toprule
&{Interaction}&{With science}&{Without science}\\
\midrule
Display     &       1.212\sym{***}&       1.726\sym{***}&       1.212\sym{***}\\
            &     (0.025)         &     (0.062)         &     (0.025)         \\
After   &       0.085\sym{***}&       0.150\sym{***}&       0.092\sym{***}\\
            &     (0.005)         &     (0.013)         &     (0.005)         \\
Display$\times$After&      -0.972\sym{***}&      -0.950\sym{***}&      -0.973\sym{***}\\
            &     (0.005)         &     (0.014)         &     (0.005)         \\
PostAge     &      -0.807\sym{***}&      -0.759\sym{***}&      -0.813\sym{***}\\
            &     (0.003)         &     (0.007)         &     (0.003)         \\
Science     &      -1.068\sym{***}&                     &                     \\
            &     (0.047)         &                     &                     \\
Display$\times$Science&       0.516\sym{***}&                     &                     \\
            &     (0.068)         &                     &                     \\
After$\times$Science&       0.115\sym{***}&                     &                     \\
            &     (0.012)         &                     &                     \\
ATT$\times$Science &       0.021         &                     &                     \\
            &     (0.015)         &                     &                     \\
Intercept      &       1.351\sym{***}&       0.326\sym{***}&       1.345\sym{***}\\
            &     (0.018)         &     (0.044)         &     (0.018)         \\
\midrule
Post-level RE & \checkmark& \checkmark& \checkmark \\
\midrule
\#Observations       &      \num{614520}         &      \num{84690}         &      \num{529830}         \\
\bottomrule
\end{tabular}}
\end{subtable}
\label{tab:did_sensitivity_post_charcs2}
\end{table}

\newpage
\section{Analysis of Deleted Posts}
\label{supp:deletedPosts}

Regression Discontinuity Design (RDD) is a quasi-experimental research design that aims to estimate the effect of a treatment by exploiting a discontinuity in the relationship between a running variable and an outcome variable at a specific threshold or cutoff point \cite{Angrist.2009,Chuai.2024}. We used note helpfulness score ($\var{NoteScore}$) as a running variable to examine the changes in the probability of post deletion within a narrow window based on a RDD model. The ``Community Notes'' program adopted several models to determine the helpfulness of community notes, while the Core model was always authoritative \cite{Chuai.2024}. To maintain consistency, we only considered note scores that were decided by the Core Model. The cut-off point of note helpfulness score was 0.4, and community notes that got score of 0.4 or above were displayed on the corresponding misleading posts. We considered the window of note helpfulness score between 0.3 and 0.5. Additionally, due to the dynamic of note writing and rating, the recalculated note helpfulness scores may be fluctuate around the cut-off point and not exactly the same with that in the production. Given this, we excluded posts that have community notes with helpfulness score between \num{0.39} and \num{0.40} (\ie [\num{0.39}, \num{0.40})). 

Subsequently, we specified our RDD model based on a logistic regression:
\begin{equation}
    \var{logit(Deletion_{i})} = \var{\beta_{0}}+\var{\beta_{1}Display_{i}}+\var{\beta_{2}NoteScore_{i}},
\end{equation}
where $\var{Display}$ indicate whether the note helpfulness score was $\geq$\num{0.40} ($=$1) or not ($=$0), and $\var{NoteScore}$ represented the running variable of the note helpfulness score. $\var{NoteScore}$ was recentered around the cut-off point. The estimation results were reported in Column (1) of \Cref{tab:rdd_deletion}. The coefficient estimate for $\var{Display}$ was \num{0.710} (99\% CI: [\num{0.588}, \num{0.832}]; $p<0.001$). We further examined the treatment effect of community notes display on the post deletion based on the odds ratio estimated for $\var{Display}$: $\var{e^{\beta}}-1$, where $\var{\beta}$ denoted the coefficient estimate of $\var{Display}$. Specifically, the estimated treatment effect was \num{1.034} (99\% CI: [\num{0.800}, \num{1.297}]; $p<0.001$). This indicated that posts with displayed community notes are 103.4\% more like to be deleted compared to posts without displayed community notes.

Additionally, we incorporated the interaction between $\var{Display}$ and $\var{NoteScore}$ in the RDD model to examine the slopes over note helpfulness scores before and after the cut-off point. The estimation results were reported in Column (2) of \Cref{tab:rdd_deletion}. The coefficient estimate of $\var{NoteScore}$ was \num{1.287} (99\% CI: [\num{-0.452}, \num{3.026}]; $p=0.057$) and not statistically significant. This suggested that the probability of post deletion kept stable over note helpfulness scores if without community notes display. Additionally, the coefficient estimate of $\var{Display} \times \var{NoteScore}$ was \num{2.875} (99\% CI: [\num{0.750}, \num{4.999}]; $p<0.001$) and significantly positive. This suggested that the efficacy of community notes in post deletion was increasing with the increase of notes helpfulness scores.

\begin{table}
\centering
\caption{Estimation results for RDD logistic regression models predicting deletion of posts. Reported are coefficient estimates with standard errors in parentheses. \sym{*} \(p<0.01\), \sym{**} \(p<0.005\), \sym{***} \(p<0.001\).}
\begin{tabular}{l*{2}{S}}
\toprule
&\multicolumn{1}{c}{(1)}           &\multicolumn{1}{c}{(2)}           \\
&\multicolumn{1}{c}{Main}&\multicolumn{1}{c}{Interaction}\\
\midrule
Display     &       0.710\sym{***}&       0.777\sym{***}\\
            &     (0.047)         &     (0.052)         \\
NoteScore   &       3.218\sym{***}&       1.287         \\
            &     (0.386)         &     (0.675)         \\
Display$\times$NoteScore&                     &       2.875\sym{***}\\
            &                     &     (0.825)         \\
Intercept   &      -2.280\sym{***}&      -2.395\sym{***}\\
            &     (0.028)         &     (0.044)         \\
\midrule
\#Posts       &      \num{81877}   &     \num{81877}\\
\bottomrule
\end{tabular}
\label{tab:rdd_deletion}
\end{table}

\end{document}